\documentclass[prl,twocolumn,groupedaddress]{revtex4}
\usepackage{graphicx,color}
\usepackage{amssymb}   
\usepackage{amsmath}
\usepackage{epstopdf}
\usepackage{natbib}
\usepackage[colorlinks,citecolor=green,urlcolor=blue,bookmarks=false,hypertexnames=true]{hyperref} 
\usepackage{bm}
\usepackage{color}
\usepackage{braket}

\begin{document}
\title{Moir\'{e} flat Chern bands and correlated quantum anomalous Hall states \\ generated by spin-orbit couplings in twisted homobilayer MoS$_2$}
\author{Benjamin T. Zhou$^1$} \thanks{Corresponding author: benjamin.zhou@ubc.ca}
\author{Shannon Egan$^1$} 
\author{Marcel Franz$^1$} \thanks{Corresponding author: franz@phas.ubc.ca}

\affiliation{$^1$Department of Physics and Astronomy \& Stewart Blusson Quantum Matter Institute,
University of British Columbia, Vancouver BC, Canada V6T 1Z4}

\begin{abstract}
We predict that in a twisted homobilayer of transition-metal dichalcogenide MoS$_2$, spin-orbit coupling in the conduction band states from $\pm K$ valleys can give rise to moir\'{e} flat bands with nonzero Chern numbers in each valley. The nontrivial band topology originates from a unique combination of angular twist and local mirror symmetry breaking in each individual layer, which results in unusual skyrmionic spin textures in momentum space with skyrmion number $\mathcal{S} = \pm 2$. Our Hartree-Fock analysis further suggests that density-density interactions generically drive the system at $1/2$-filling into a valley-polarized state, which realizes a correlated quantum anomalous Hall state with Chern number $\mathcal{C} = \pm 2$. Effects of displacement fields are discussed with comparison to nontrivial topology from layer-pseudospin magnetic fields. 
\end{abstract}

\maketitle

\emph{Introduction.}--- The discovery of possible correlated insulating states \cite{Cao1} and superconductivity \cite{Cao2} in magic-angle twisted bilayer graphene has paved a new avenue toward engineering electronic structures where interactions play a decisive role \cite{Carr, MacDonald1, MacDonald2, Adrian1, Noah, Koshino, Isobe, Wu1, Vafek1, Yang, Biao, Gonzalez, Xie, Andrei, Efetov, Perge, Yazdani1}. Lately, nontrivial topological properties brought about by the moir\'{e} patterns were also unveiled \cite{Jianpeng1, Song}. Under appropriate symmetry breaking conditions, flat bands in twisted graphene acquire nonzero Chern numbers \cite{Zaletel, Senthil1, Senthil2, Wenyu}, which manifest experimentally as quantum anomalous Hall (QAH) states when spin or valley degeneracies are lifted spontaneously through electronic correlations \cite{Sharpe, Young, Yazdani2}. The interplay between correlation and topology uncovered in these systems indicates possibilities for novel topological phases beyond conventional non-interacting theories. 

Motivated by advances in twisted graphene systems, explorations into moir\'{e} superlattices formed by other two-dimensional materials, such as transition-metal dichalcogenides (TMDs) \cite{Wu2, Wu3, LeRoy, Fai1, Wang, Dean, Bi} and copper oxides \cite{Marcel, Pixley}, have also seen rapid progress recently. In particular, moir\'{e} flat bands in a twisted homobilayer TMD formed by the valence band (VB) states in each $K$-valley were shown to carry nonzero Chern numbers \cite{Wu2}. The nontrivial band topology arises from combined effects of moir\'{e} potential and interlayer coupling, which act together as a layer-pseudospin magnetic field and create skyrmionic pseudospin textures in superlattice cells. The strong spin-valley locking due to giant Ising-type spin-orbit coupling (SOC) of order $\sim$ 100 meVs \cite{Xiao, GuiBin} further entails a possible quantum spin Hall state. 

\begin{figure}
\centering
\includegraphics[width=3.5in]{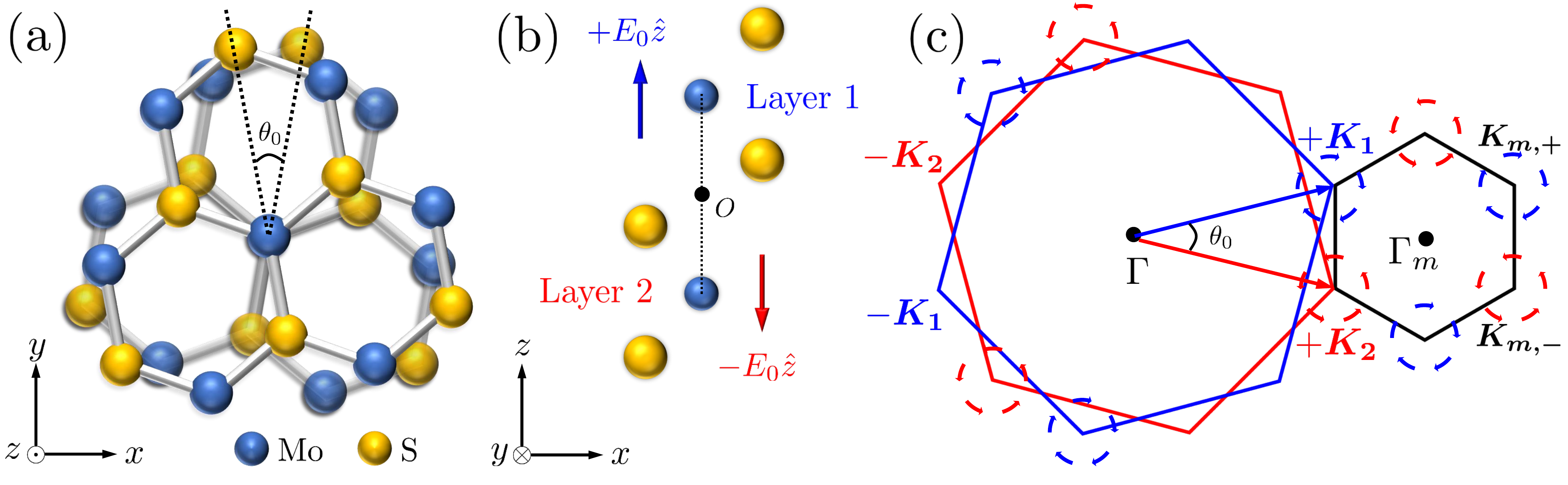}
\caption{Schematic crystal structure of twisted bilayer MoS$_2$ with (a) top view and (b) side view. Electrons in different layers experience opposite effective electric fields due to local lack of mirror symmetry. (c) Moir\'{e} Brillouin zone(MBZ) formed by twisting hexagonal Brillouin zones of each layer. The in-plane spinor field exhibits opposite vorticities at $K_{m,+}$ and $K_{m,-}$ of MBZ due to layer-dependent Rashba effects.}
\label{FIG01}
\end{figure}

While flat bands from VB states have enjoyed wide interest, moir\'{e} physics arising from conduction band (CB) states in twisted TMDs remains largely unexplored. One possible complication lies in the relatively weak SOC in the conduction bands on the scale of a few to tens of meVs \cite{GuiBin, Rossier}, which may cause crossings between spin-up and spin-down moir\'{e} bands. Recent works on untwisted TMDs, on the other hand, revealed that the small spin-orbit splitting in CB states, when combined with Rashba SOC introduced by mirror symmetry breaking, can lead to nontrivial Berry phase effects \cite{Benjamin, Taguchi, Lee}. Notably, upon assembling two identical TMD layers into a twisted bilayer, the mirror symmetry is broken locally in each layer (Fig.\ref{FIG01}(a)-(b)) and Rashba SOC is expected to arise. How this SOC would affect the band topology of twisted TMDs is an outstanding question.

In this Letter we establish a novel topological phase generated by SOC in CB moir\'{e} bands in twisted TMDs, which is essentially different from those in twisted graphene \cite{Jianpeng1, Song, Zaletel, Senthil1, Senthil2, Wenyu} where SOC is negligible, and those in the VB moir\'{e} bands of twisted TMDs where SOC plays no essential role in creating nonzero Chern numbers \cite{Wu2}. Focusing on the specific case of a twisted homobilayer MoS$_2$ we predict that an interplay among angular twist, local symmetry breaking and spin-orbit coupling in the CB states creates moir\'{e} flat bands with nonzero Chern numbers $\mathcal{C} = \pm 2$, and density-density interactions generically drive the system at $1/2$-filling into a valley-polarized correlated QAH state.
 We further show that the Chern bands generated by SOC stay robust against displacement fields which would otherwise destroy the nontrivial topology from layer-pseudospins \cite{Wu2}. 

\emph{Continuum model for twisted homobilayer MoS$_2$.}--- Stacking two monolayers of MoS$_2$ with a small relative twist angle $\theta_0$ results in a moir\'{e} superlattice with lattice constant $a_M = a/\theta_0$, where $a$ is the lattice constant of each monolayer. The system has a symmetry group of $D_3 \otimes \{U_{v}(1), \mathcal{T}\}$, where $U_{v}(1)$ stands for valley conservation, $\mathcal{T}$ for time-reversal symmetry, and the point group $D_3$ is generated by the three-fold rotation about the $z$-axis ($C_{3z}$) and a two-fold rotation about the $y$-axis ($C_{2y}$) (Fig.\ref{FIG01}(a)). For an aligned bilayer at $\theta_0 = 0$, the mirror symmetry $\mathcal{M}_z$ about the horizontal 2D mirror plane, which is respected in a monolayer, is broken locally on each layer. This generates uniform electric fields of opposite signs in different layers, which remain effective upon a small angular twist with $\theta_0 \sim 1^{\circ}$ (Fig.\ref{FIG01}(b)). Based on models of monolayer MoS$_2$ with broken $\mathcal{M}_z$ \cite{Benjamin, Taguchi, Lee}, the effective Hamiltonian for CB minima at $\pm K$ valleys of two twisted isolated layers can be written as:
\begin{eqnarray} \label{eq:Hisolated}
\mathcal{H}^{\xi}_{0, l} &=& \sum_{\bm{k}, \alpha\beta} c^{\dagger}_{\bm{k},  l, \alpha} h^{\xi}_{l, \alpha\beta} (\bm{k}) c_{\bm{k}, l, \beta}, \\\nonumber
   h^{\xi}_{l} (\bm{k}) &=& \frac{\hbar^2 (\bm{k} - \xi \bm{K}_{l})^2}{2 m^{\ast}} - \mu + \xi \beta_{\rm so} \sigma_z \\\nonumber
   &+& (-1)^l \alpha_{\rm so} [(\bm{k} - \xi \bm{K}_{l}) \times \bm{\sigma}] \cdot \hat{z}, 
\end{eqnarray}
where $\xi = \pm$ denotes the valley index, $l = 1 (2)$ labels the top (bottom) layer, $\alpha, \beta$ label the spin indices with the $\sigma$-matrices representing the usual Pauli matrices for spins. $\bm{K}_{1} \equiv \bm{K}_{m,+}$ and $\bm{K}_{2}\equiv \bm{K}_{m,-}$ are the shifted $+\bm{K}$-points on layer 1 and layer 2 (Fig.\ref{FIG01}(c)) due to the angular twist of $\theta_0$ equivalent to rotating the top (bottom) layer by an angle $\theta_0/2$ ($-\theta_0/2$) about the $z$-axis. In Eq.\ref{eq:Hisolated}, $m^{\ast} \approx 0.5 m_e$ is the CB effective mass at $K$, $\mu$ is the Fermi energy measured from band minima. The $\alpha_{\rm so}$-term and $\beta_{\rm so}$-term account for the Rashba SOC and the Ising SOC, respectively. The Ising SOC has a valley-dependent sign due to time-reversal symmetry and causes a splitting of $\Delta_{\rm so} = 2|\beta_{\rm so}|$ in the spin-up and spin-down CB states. The Rashba SOC has a layer-dependent sign imposed by $C_{2y}$ which swaps the two layers. Spatial variations of $\alpha_{\rm so}$ are discussed in Supplemental Material (SM). 

\begin{figure}
\centering
\includegraphics[width=3.4in]{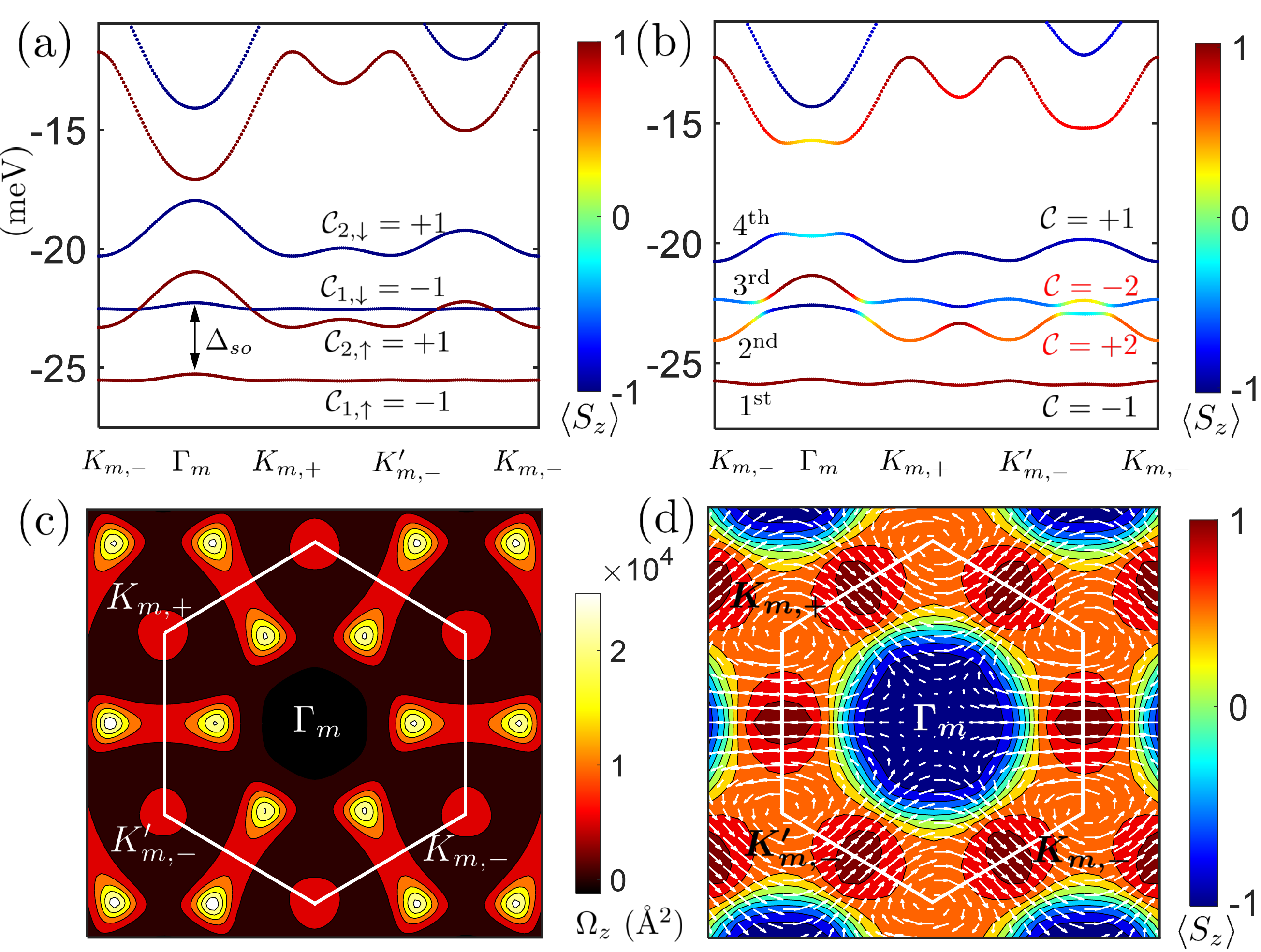}
\caption{Moir\'{e} bands of valley $\xi = +$ formed by CB states of twisted homobilayer MoS$_2$ at $\theta_0 = 1.4^{\circ}$ with (a) $\alpha_{\rm so} = 0$ and (b) $\alpha_{\rm so} \neq 0$. The colorbars indicate the out-of-plane spin expectation value $\braket{S_z}$ in units of $\hbar/2$. (c) Momentum-space profile of Berry curvature $\Omega_z$ in units of {\AA}$^2$ in the 2$^{\text{nd}}$ moir\'{e} band in (b). Large $\Omega_z$ of order 10$^4${\AA}$^2$ emerges as a result of the nontrivial gap induced by SOC. (d) Momentum-space spin texture of the 2$^{\text{nd}}$ moir\'{e} band in (b). White arrows indicate the in-plane spinor field, which has opposite vorticities at $K_{m,+}$ and $K_{m,-}$ and leads to a nontrivial skyrmion number $\mathcal{S} = +2$. Parameters used are presented in Table \ref{table:01}.}
\label{FIG02}
\end{figure}

Spatial modulations due to the formation of moir\'{e} pattern can be captured by introducing coupling terms between states at $\bm{k}$ and $\bm{k} +\bm{G}_{M,j}$ \cite{Wu2, Senthil1}, where $\bm{G}_{M,j} = -\frac{4\pi}{\sqrt{3} a_M} (\cos \frac{(j-1)\pi}{3} , \sin \frac{(j-1)\pi}{3})$ are the reciprocal vectors of the moir\'{e} superlattice. The momentum-space moir\'{e} potential is given by
\begin{eqnarray} \label{eq:moirepotential}
\mathcal{H}_M = \sum_{\bm{k}} \sum_{j=1}^6 \sum_{l=1,2} \sum_{\alpha = \uparrow, \downarrow} c^{\dagger}_{\bm{k}, l,\alpha} V_{l, \bm{G}_{M,j}} c_{\bm{k}+\bm{G}_{M,j}, l,\alpha},
\end{eqnarray}
where $V_{l, \bm{G}_{M,j}} =  V_M (\cos\psi - i (-1)^j l \sin\psi )$ are complex coupling parameters with amplitude $V_M$ and phase $\psi$.

The effective inter-layer tunneling for states at $K$-valleys is modeled following the general recipe of two-center approximation \cite{MacDonald1, Wu2}, which can be written as
\begin{eqnarray}\label{eq:interlayerH}
\mathcal{H}^{\xi}_{T} 
= -w_0 \sum_{\bm{k}}  c^{\dagger}_{\bm{k}, 1} ( c_{\bm{k}, 2} +  c_{\bm{k}+\xi\bm{G}_{M,2}, 2} 
+  c_{\bm{k}+\xi\bm{G}_{M,3}, 2} ) + {\rm h.c.},
\end{eqnarray}
where $w_0$ denotes the effective inter-layer tunneling amplitude near $K$, which is extrapolated to be $w_{0} \approx 10$ meV using a combined approach of empirical models and Slater-Koster methods (see the Supplemental Material \cite{SM} for details). The total non-interacting continuum model thus reads: $\mathcal{H}_0 = \sum_{\xi = \pm} \sum_{l=1,2} \mathcal{H}^{\xi}_{0,l} + \mathcal{H}^{\xi}_{T}  + \mathcal{H}_{M}$ with parameters tabulated in Table \ref{table:01}. 

\emph{Moir\'{e} flat Chern bands and skyrmionic spin textures.}--- From general considerations, level spacings among the lowest moir\'{e} bands correspond roughly to the quantization energy $\Delta E \sim \hbar^2 \pi^2/(2m^{*} a_M^2) $ of an electron confined in a superlattice cell with size $a_M^2 =a^2/\theta^2_0$. For $\theta_0 \sim 1^{\circ} - 2^{\circ}$, $\Delta E \sim 1-10$ meV in twisted MoS$_2$, which is comparable to the spin-orbit splitting $\Delta_{\rm so} \equiv 2|\beta_{\rm so}|= 3$ meV caused by the Ising SOC \cite{GuiBin}. If spin is conserved (\textit{e.g.}, in the absence of Rashba SOC), spin-up and spin-down moir\'{e} bands are expected to cross in general. 

By setting $\alpha_{\rm so}=0$ in $\mathcal{H}_0$, we find that for $\xi = +$, the 1$^{\text{st}}$ spin-down and the 2$^{\text{nd}}$ spin-up moir\'{e} bands cross each other for $\theta_0 \in (1.25^{\circ},1.5^{\circ})$. As a specific example, spin-resolved moir\'{e} bands at $\theta_0 = 1.4^{\circ}$ with $\alpha_{\rm so} = 0$ are shown in Fig.\ref{FIG02}(a). With spin conservation in this case, the spin Chern number is well-defined, and the spin-up and spin-down bands involved in the level crossing have Chern numbers ($\mathcal{C}_{1, \downarrow} = +1$, $\mathcal{C}_{2, \uparrow} = -1$), which originate from the layer-pseudospin magnetic fields as in moir\'{e} bands from VB states \cite{Wu2}. Upon turning on $\alpha_{\rm so}$, crossing points are gapped out by the layer-dependent Rashba SOC which mix the spin-up and spin-down species (Fig.\ref{FIG02}(b)). As a result of this mixing, the original $\mathcal{C}_{1, \downarrow} $ and $\mathcal{C}_{2, \uparrow}$ from layer pseudospins annihilate each other; meanwhile, a new pair of flat bands (2$^{\text{nd}}$ and 3$^{\text{rd}}$ bands in Fig.\ref{FIG02}(b)) with Chern numbers $\mathcal{C} = \pm 2$ emerge (see SM \cite{SM} for details on Chern number calculations). The nontrivial gap induced by Rashba SOCs is further signified by giant Berry curvatures of order $10^4${\AA}$^2$ in the MBZ (Fig.\ref{FIG02}(c)). 

The unusual Chern numbers $\mathcal{C} = \pm 2$ arise from a \textit{unique} combination of angular twist and local mirror-symmetry breaking: as moir\'{e} bands at opposite $\bm{K}_{m,+}$ and $\bm{K}_{m,-}$ points originate from states in different layers (Fig.\ref{FIG01}(c)), the layer-dependent Rashba SOCs (Eq.\ref{eq:Hisolated}) create an in-plane spinor field with opposite vorticities at $\bm{K}_{m,+}$ and $\bm{K}_{m,-}$ in the MBZ. This unsual pattern causes the in-plane spin to wind twice as one goes around a loop enclosing the $\Gamma_m$-point in the MBZ (Fig.\ref{FIG02}(d)). As we confirm numerically, the spin textures of the 2$^{\textrm{nd}}$ and 3$^{\textrm{rd}}$ moir\'{e} bands are characterized by nonzero skyrmion numbers $\mathcal{S} = \int d^2 \bm{k} \hat{n}_{\bm{k}} \cdot (\partial_{k_x} \hat{n}_{\bm{k}} \times \partial_{k_y} \hat{n}_{\bm{k}} ) / 4\pi = \pm 2$ ($\hat{n}_{\bm{k}}$: spin orientation of state at $\bm{k}$), and these two bands describe a two-level system with one-to-one correspondence between Chern numbers $\mathcal{C}$ and skyrmion number $\mathcal{S}$ \cite{ Bernevig}.

\begin{table}[tp] 
\caption{Parameters used in $\mathcal{H}_0$ for twisted homobilayer MoS$_2$ with monolayer lattice constant $a = 3.16 {\AA}$.}
\centering
\begin{tabular}{c c c c c c c}
\hline \hline
 $\mu$ & $\alpha_{\rm so}$ &  $\beta_{\rm so}$ & $V_M$ &  $\psi$  &  $w_0$ \tabularnewline
\hline 
 0 meV & 80 meV$\cdot{\AA}$  & -1.5 meV & 10 meV & -89.6$^{\circ}$ & 10 meV \tabularnewline
\hline\hline
\end{tabular}
\label{table:01}
\end{table}

\emph{Correlated QAH state at $1/2$-filling.}--- Given the narrow bandwidth $W \approx 1$ meV of moir\'{e} Chern bands (Fig.\ref{FIG02}(b)), the characteristic Coulomb energy scale $g_C \simeq e^2/\epsilon a_M \approx 10$ meV overwhelms the kinetic energy and correlation physics is expected to arise (assuming $\theta_0 = 1.4^{\circ}$ and dielectric constant $\epsilon = 10$ \cite{Bi}). Motivated by the observation that twisted graphene at $3/4-$filling can behave as an SU(4)-quantum Hall ferromagnet \cite{Senthil1, Sharpe, Young, Yazdani2}, and the fact that all CB moir\'{e} bands in Fig.\ref{FIG02}(b) involves only a two-fold valley degeneracy (note that spin SU(2)-symmetry is already broken by SOC), we examine correlation effects when the 2$^{\text{nd}}$ moir\'{e} band is half-filled. For simplicity, we drop the band index $n$ in the following.

Including density-density interactions, the low-energy effective Hamiltonian: $\mathcal{H}_{\rm eff} = \mathcal{H}_{0, \rm eff} + \mathcal{H}_{I, \rm eff}$, where $
\mathcal{H}_{0, \rm eff} = \sum_{\bm{k}, \xi = \pm} E_{\xi} (\bm{k}) c^{\dagger}_{\xi} (\bm{k}) c_ {\xi} (\bm{k}) $
is the non-interacting part with band energies $E_{\xi}(\bm{k})$, and the interacting Hamiltonian compatible with $U_{v}(1)$ and $\mathcal{T}$ reads
\begin{eqnarray}
\mathcal{H}_{I, \rm eff} &=& \sum_{\bm{k}\bm{k}'\bm{q}, \xi\xi'} \frac{V(\bm{q})}{A} \Lambda^{\xi}  (\bm{k}+\bm{q}, \bm{k}) \Lambda^{\xi'} (\bm{k}'-\bm{q} , \bm{k}') \\\nonumber
&\times &c^{\dagger}_{\xi}  (\bm{k}+\bm{q}) c^{\dagger}_{\xi'}  (\bm{k}'-\bm{q}) c_{\xi'}  (\bm{k}') c_{\xi}  (\bm{k}),
\end{eqnarray}
where $V(\bm{q})$ is the Fourier transform of the two-body interaction, $A$ is the total area of the system, and $\Lambda^{\xi}  (\bm{k}+\bm{q}, \bm{k}) \equiv \braket{u_{\xi, \bm{k}+\bm{q}} | u_{\xi, \bm{k}} }$ ($\ket{u_{\xi, \bm{k}}}$: periodic part of the Bloch states) are the form factors arising from projecting the density-density interaction to active bands. The general trial Hartree-Fock ground state at $1/2$-filling has the form: $\ket{\Phi} = \prod _{\bm{k}\in \text{MBZ}} [w_{+}(\bm{k}) c^{\dagger}_{+} (\bm{k}) +  w_{-}(\bm{k}) c^{\dagger}_{-} (\bm{k})]  \ket{\Phi_0}$,
where $\ket{\Phi_0}$ is the vaccuum state with all lower-lying bands filled, and $w_{\xi}(\bm{k})$ are the variational parameters. 

\begin{figure}
\centering
\includegraphics[width=3.4in]{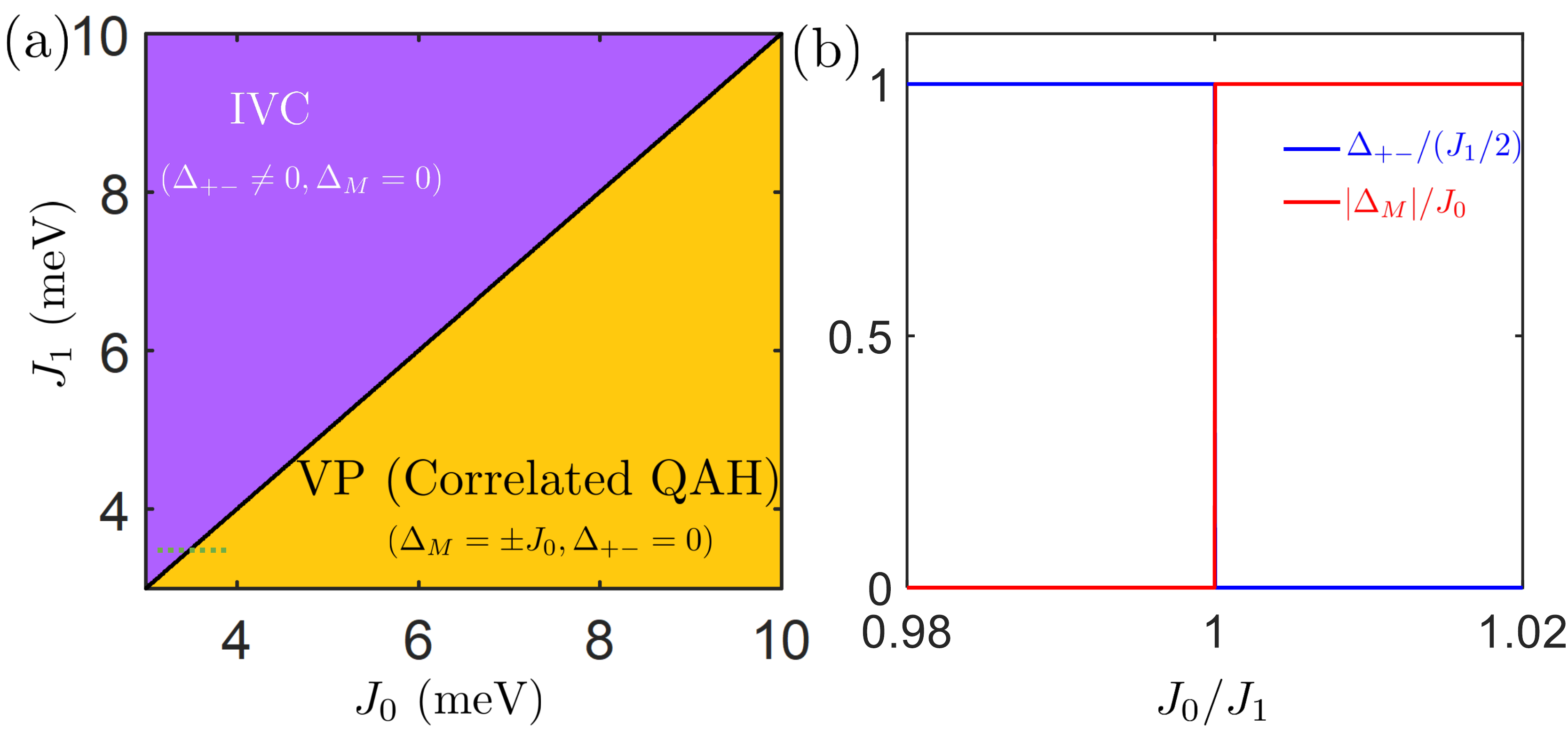}
\caption{(a) $J_0- J_1$ phase diagram of the system with the 2$^{\text{nd}}$ moir\'{e} band at $1/2$-filling. Under repulsive density-density interactions, only the $J_1 \leq J_0$ regime can be accessed \cite{SM}, where the valley-polarized(VP) state is generically favored and the system becomes a correlated QAH insulator. (b) Evolutions of the order parameters $\Delta_{+-}$ and $\Delta_M \equiv \Delta_+ - \Delta_-$ as the ratio $J_0/J_1$ is tuned along the line-cut (green dashed line in (a)) across the phase boundary. A first-order transition happens at the critical point $J_0 \simeq J_1$.}
\label{FIG03}
\end{figure}

Minimizing $E^{\Phi} \equiv \braket{\Phi|\mathcal{H}_{\rm eff}|\Phi}$ with mean-field approximations for Fock exchange terms (see SM \cite{SM}) leads to a set of self-consistent equations: 
\begin{equation}\label{eq:MFEqn}
\begin{cases}
\Delta_{+} + \Delta_{-} = J_0, \\
\Delta_{+} - \Delta_{-} = \frac{J_0}{2N} \sum_{\bm{k}}  \frac{(\Delta_{+} - \Delta_{-}) - \delta(\bm{k})}{D(\bm{k})}, \\
\Delta_{+-} = \frac{J_1}{2N} \sum_{\bm{k}}  \frac{\Delta_{+-}}{D(\bm{k})}.
\end{cases}
\end{equation}
Here, $J_0$ and $J_1$ denote the mean-field \textit{intra-valley} and \textit{inter-valley} exchange coupling constants, $\Delta_{\xi} = J_0 \bar{n}_{\xi}$ ($\bar{n}_{\xi}$: mean occupancy for valley $\xi$) and $\Delta_{+-}$ are the \textit{intra-valley} and \textit{inter-valley} order parameters, which characterize the average energy gains from \textit{intra-valley} and \textit{inter-valley} exchange interactions, respectively. $\delta(\bm{k}) \equiv E_{+}(\bm{k}) - E_{-}(\bm{k})$ denotes energy difference between bands from two valleys $\xi=\pm$, $D(\bm{k}) = \sqrt{\Delta^2_{+-} + [\delta(\bm{k}) - \Delta_M]^2/4}$, where we introduce the \textit{valley magnetic order} $\Delta_M \equiv \Delta_{+} - \Delta_{-} = J_0 (\bar{n}_{+} - \bar{n}_{-})$. 

Solutions of Eq.\ref{eq:MFEqn} are divided into two classes: (i) the valley-polarized (VP) state \cite{Adrian1, Senthil1}, in which one out of the two nearly degenerate moir\'{e} bands from two valleys is fully filled, with spontaneous $\mathcal{T}$-symmetry breaking: $\Delta_{M} = \xi J_0, \Delta_{+-} = 0, w_{\xi}(\bm{k}) = 1$; (ii) the inter-valley coherent (IVC) state, in which the Slater determinant is formed by coherent superpositions of states from the two valleys, with spontaneous U$_{v}(1)$-symmetry breaking \cite{Adrian1, Senthil1, Senthil2}: $\Delta_M = 0, \Delta_{+-} \approx J_1/2$, $w_{+,-}(\bm{k}) \neq 0$. By comparing the energies of the VP and IVC states, we obtain the mean-field $J_0 - J_1$ phase diagram for the range $0.3 g_C \leq J_0, J_1 \leq g_C$ (Fig.\ref{FIG03}(a)). The IVC state is favored throughout the entire $J_1 \geq J_0$ regime, while the VP phase takes up most of the $J_1 < J_0$ regime. By tuning $J_0/J_1$ across the phase boundary (green dashed line in Fig.\ref{FIG03}(a)), a first-order transition occurs at $J_0 \simeq J_1$ (Fig.\ref{FIG03}(b)).

It is worth noting that given a general repulsive interaction $V(\bm{q}) > 0$, the inequality $2|\Lambda^{+}(\bm{k}', \bm{k})|| \Lambda^{-}(\bm{k}, \bm{k}') | \leq |\Lambda^{+}(\bm{k}', \bm{k})|^2 + |\Lambda^{-}(\bm{k}, \bm{k}')|^2$ together with $\mathcal{T}$-symmetry leads to $J_0 \geq J_1$, regardless of details of the form factors and the microscopic interaction \cite{SM}. 
Thus, electron correlations most likely drive the system at $1/2$-filling into a $\mathcal{T}$-broken VP state, which realizes a correlated QAH state with $\mathcal{C} = \pm 2$. The case of general filling away from 1/2 is discussed in SM \cite{SM}.

\emph{Effects of displacement fields and comparison with Chern bands from layer pseudospins.}--- As shown in Fig.\ref{FIG02}(a), with $\alpha_{\rm so} = 0$, nonzero Chern numbers also arise in CB moir\'{e} bands due to layer pseudospin magnetic fields, with the same origin as those in VB moir\'{e} bands \cite{Wu2}. Thus, two different mechanisms for moir\'{e} Chern bands, one from layer pseudospin and the other from SOCs, \textit{coexist} in the CB moir\'{e} bands. In contrast, the Rashba SOC is unimportant in the VB moir\'{e} bands where spin-up and spin-down bands are separated by $100-400$ meV \cite{GuiBin}, and the VB moir\'{e} band topology is governed by layer pseudospin alone.

The nontrivial topology from layer pseudopsin is shown to be prone to displacement fields which can destroy the skyrmionic pseudospin textures by polarizing the layer-pseudospins \cite{Wu2}. On the other hand, the SOC mechanism has an independent origin in the avoided level crossings between spin-up and spin-down bands and does not require nontrivial layer pseudospin textures. Thus one would expect the nontrivial band topology in the CB moiré bands to be more robust against displacement fields than its VB counterpart.

\begin{figure}
\centering
\includegraphics[width=3.4in]{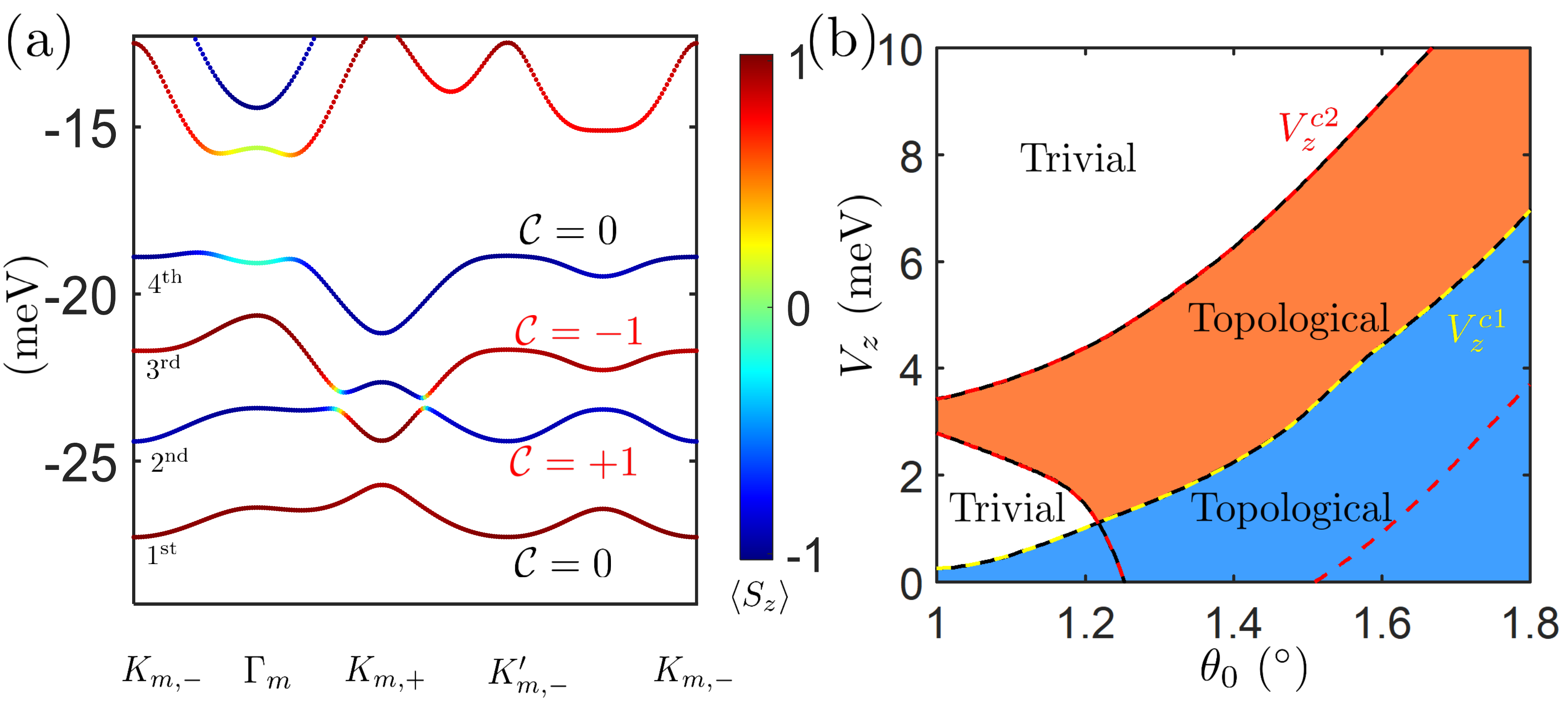}
\caption{(a) CB Moir\'{e} bands for valley $\xi = +$ of twisted bilayer MoS$_2$ at $\theta_0 = 1.4^{\circ}$ and $V_z = 4$ meV. Under finite $V_z$, spin-up and spin-down bands still cross, and the avoided level crossing due to SOC result in a pair of bands with $\mathcal{C} = \pm 1$. (b) Topological phase diagram of CB moir\'{e} bands as a function of $\theta_0$ and $V_z$, with an upper boundary $V_z^{c2}$ of the entire topological regime. Yellow dashed line: critical $V_z^{c1}$ above which nontrivial topology generated by layer-pseudospin is destroyed. Values of $V^{c1}_z$ are similar to those found in VB moir\'{e} bands \cite{Wu2}. Area enclosed by the red dashed lines: regime with level crossings between spin-up and spin-down CB moir\'{e} bands when $\alpha_{\rm so} = 0$. Area in orange: topological regime where layer pseudospin mechanism in Ref.\cite{Wu2} fails. }
\label{FIG04}
\end{figure}

To demonstrate the effects of displacement fields, we follow Refs.\cite{Wu2} and introduce a layer-dependent potential $V_{D, l} = (-1)^l V_z/2, (l=1,2)$ in $\mathcal{H}_0$. The moir\'{e} bands at $\theta_0 = 1.4^{\circ}$ under finite $V_z$ are shown in Fig.\ref{FIG04}(a), where we set $V_z = 4$ meV strong enough to destroy the nontrivial topology generated by layer pseudospins (Fig.\ref{FIG04}(b)). Clearly, the avoided level crossings still occur and a pair of Chern bands with Chern numbers $(+1, -1)$ remain. The Chern number is reduced from $\mathcal{C} = \pm 2$ to $\mathcal{C} = \pm 1$ under $V_z$ because the displacement field drives a band inversion near $\bm{K}_{m,+}$ and mediates a Chern number exchange $\Delta \mathcal{C} = \pm 1$ between the 1$^{\text{st}}$ and 2$^{\text{nd}}$ moir\'{e} bands.

The topological phase diagram for CB moir\'{e} bands as a function of $\theta_0$ and $V_z$ is shown in Fig.\ref{FIG04}(b). Due to the extra mechanism from SOCs, the critical displacement field for the nontrivial topological regime is enhanced to be $V^{c2}_z$, which is 2-4 times of the critical $V^{c1}_z$ (yellow dashed line) needed to destroy the layer pseudospin mechanism for $\theta_0 \sim 1^{\circ}$ ($V^{c1}_z$ obtained by setting $\alpha_{\rm so} = 0$ in $\mathcal{H}_0$). Importantly, there is a wide parameter regime in the phase diagram (depicted in orange) where the layer pseudospin mechanism proposed in Ref.\cite{Wu2} fails while the CB moir\'{e} bands remain topological, which confirms that the CB states are more robust than VB states against displacement fields. 

\emph{Conclusion and Discussions. }--- While on-going activities in twisted TMDs have focused mainly on VB moir\'{e} bands  \cite{Wu2, Wu3, LeRoy, Fai1, Wang, Dean, Bi}, our proposal of moir\'{e} Chern bands generated by SOC opens a new pathway into the largely unknown territory of CB moir\'{e} physics. When these Chern bands are half-filled, according to our predictions, electron interactions can lead to a spontaneously $\mathcal{T}$-broken VP phase and realize a correlated QAH state, which will manifest itself through quantized Hall conductance in transport measurements. 

To realize the topological moir\'{e} bands generated by SOC, any finite Rashba SOC which is allowed by the $D_3$-symmetry of the system would be sufficient, in principle, to induce a nontrivial gap. Notably, as MoS$_2$ is intrinsically semiconducting (with Fermi level $\sim 0.8$ eV below the conduction band edge) \cite{Xiao, GuiBin, Fai2}, it is necessary to gate the chemical potential such that the CB moir\'{e} bands are filled in the first place. With the dual-gate setup used widely in experiments \cite{Cao1, Dean}, local mirror symmetry breaking can be further enhanced by interfacial contact with gating electrodes, which enhances the SOC gap in the non-interacting bands. Due to the correlated nature of the QAH state at 1/2-filling, the actual topological gap to be manifested experimentally is not solely determined by the SOC gap in the non-interacting bands; instead it should be largely governed by the intra-valley exchange coupling $J_0 \sim 10$ meV for $\theta_0 \sim 1^{\circ}$ \cite{SM}. This sizable correlation-induced topological gap will reduce complications of disorder and thermal effects in experimental detection of the predicted correlated QAH state. However, as $\theta_0$ increases, correlation effects become weaker and the moir\'{e} bands more dispersive, thus the predicted correlated QAH phase would be less robust in the large twist angle regime.\\

\emph{Acknowledgement.} --- B.T.Z. thanks Zefei Wu and Ziliang Ye for inspiring discussions on experimental aspects of twisted MoS$_2$. This work was supported by NSERC and the Canada First Research Excellence Fund, Quantum Materials and Future Technologies Program. B.T.Z. further acknowledges the support of the Croucher Foundation.

\clearpage

\onecolumngrid
\begin{center}
\textbf{\large Supplemental Material for\\ ``Moir\'{e} Chern bands and correlated quantum anomalous Hall states\\ generated by spin-orbit couplings in twisted homobilayer MoS$_2$''} \\[.2cm]
Benjamin T. Zhou,$^{1}$ Shannon Egan,$^{1}$ Marcel Franz$^{1}$\\[.1cm]
{\itshape ${}^1$Department of Physics and Astronomy \& Stewart Blusson Quantum Matter Institute,
University of British Columbia, Vancouver BC, Canada V6T 1Z4}	\\
\end{center}
\setcounter{equation}{0}
\setcounter{section}{0}
\setcounter{figure}{0}
\setcounter{table}{0}
\setcounter{page}{1}
\renewcommand{\theequation}{S\arabic{equation}}
\renewcommand{\thetable}{S\arabic{table}}
\renewcommand{\thesection}{\arabic{section}}
\renewcommand{\thefigure}{S\arabic{figure}}
\renewcommand{\bibnumfmt}[1]{[S#1]}
\renewcommand{\citenumfont}[1]{S#1}
\makeatletter

\section*{I. Calculation of effective inter-layer tunneling parameter $w_0$}

In the main text, we mention that an effective inter-layer tunneling parameter $w_0$ is extrapolated using a combined approach of empirical model and Slater-Koster method. Here, we present details on the calculation of $w_0$.

The inter-layer tunneling between two MoS$_2$ monolayers with a small twist angle can be conveniently modeled by the two-center approximation following the general recipe introduced in previous studies on twisted bilayer graphene \cite{MacDonaldS} and twisted homobilayer MoTe$_2$ \cite{WuS}. Particularly, electronic states near the conduction band minimum of monolayer transition-metal dichalcogenides(TMDs) are predominantly from the $4 d_{z^2}$-orbitals of the transition metal atoms \cite{GuiBinS} and the inter-layer tunneling is expected to be mediated by the $\sigma$-bonding between $4 d_{z^2}$-orbitals in each layer. To start with, we consider fermionic operators that create Bloch states of the form:
\begin{eqnarray}
c^{\dagger}_{\bm{k}, 1, \alpha} &=& \frac{1}{\sqrt{N}} \sum_{\bm{R}} e^{i \bm{k} \cdot \bm{R} } c^{\dagger}_{1, \alpha} (\bm{R}), \\\nonumber
c^{\dagger}_{\bm{k}', 2, \beta} &=& \frac{1}{\sqrt{N}} \sum_{\bm{R}'} e^{i \bm{k}' \cdot (\bm{R}' + \bm{d}_0)} c^{\dagger}_{2, \beta} (\bm{R}'+ \bm{d}_0).
\end{eqnarray}
Here, $\bm{k}, \bm{R}$ and $\bm{k}', \bm{R}'$ denote the momentum and lattice vectors of layer 1 and 2, respectively. $c^{\dagger}_{l, \alpha} (\bm{R})$ creates a localized Wannier orbital of $ d_{z^2}$ character with spin $\alpha$ at site $\bm{R}$ in layer $l$. In general, the inter-layer tunneling Hamiltonian is given by:
\begin{eqnarray}
\mathcal{H}_{T} = \sum_{\bm{k},\bm{k}'} \sum_{\alpha, \beta} c^{\dagger}_{\bm{k}, 1, \alpha} T_{\alpha\beta} (\bm{k},\bm{k}') c_{\bm{k}', 2, \beta} + h.c.
\end{eqnarray}
where the tunneling matrix $T$ reads:
\begin{eqnarray}
T_{\alpha\beta}(\bm{k},\bm{k}') &=& \braket{\bm{k}, 1,\alpha| \mathcal{H}_{T} | \bm{k}', 2, \beta} \\\nonumber
&=& \frac{1}{N} \sum_{\bm{R},\bm{R}'} e^{-i \bm{k} \cdot \bm{R} } e^{i \bm{k}' \cdot (\bm{R}' + \bm{d}_0) } \braket{\bm{R}, 1,\alpha| \mathcal{H}_{T} | \bm{R}' + \bm{d}_0 , 2, \beta} \\\nonumber
(\text{two-center approximation) }&=&  -\frac{1}{N} \sum_{\bm{R},\bm{R}'} e^{-i \bm{k} \cdot \bm{R} } e^{i \bm{k}' \cdot (\bm{R}' + \bm{d}_0) } t_{\alpha\beta} (\bm{R} - \bm{R}' -\bm{d}_0) \\\nonumber
&=&  -\frac{1}{N} \sum_{\bm{R},\bm{R}',\bm{q}} e^{-i \bm{k} \cdot \bm{R} } e^{i \bm{k}' \cdot (\bm{R}' + \bm{d}_0) } \frac{t_{\alpha\beta} (\bm{q})}{A} e^{i \bm{q}\cdot (\bm{R} - \bm{R}' -\bm{d}_0) } \\\nonumber
&=&  -\frac{1}{N} \sum_{\bm{q}} \frac{t_{\alpha\beta} (\bm{q})}{A}e^{i (\bm{k}' -\bm{q}) \cdot \bm{d}_0} \big(\sum_{\bm{R}} e^{i (\bm{q}-\bm{k}) \cdot \bm{R} }\big)  \big(\sum_{\bm{R}'} e^{i (\bm{k}' -\bm{q}) \cdot \bm{R}'  } \big) \\\nonumber
&=&  -\sum_{\bm{G}, \bm{G}'} \frac{t_{\alpha\beta} (\bm{k}+\bm{G})}{\Omega} e^{-i \bm{G}'  \cdot \bm{d}_0} \delta_{\bm{k}+\bm{G}, \bm{k}'+\bm{G}'}
\end{eqnarray}
Here, $t_{\alpha\beta}(\bm{q}) = \int d^2\bm{r} t_{\alpha\beta}(\bm{r}) e^{-i\bm{q}\cdot\bm{r}}$ is the Fourier transform of $t_{\alpha\beta}(\bm{r})$, $A$ is the total area of the system, $\Omega = A/N$ is the area of the primitive unit cell. In the last step of the derivation above, we made use of the identity $\sum_{\bm{R}} e^{i (\bm{q}-\bm{k}) \cdot \bm{R} } =N \sum_{\bm{G}} \delta_{\bm{q}, \bm{k}+\bm{G}}$. 

To simplify the expression of $T_{\alpha\beta}(\bm{k},\bm{k}')$ above, we note that for the relevant electronic states near the $K$-points with $\bm{k} \sim \xi\bm{K}$, the dominant $t_{\alpha\beta}(\bm{q})$-terms that enter the continuum model for valley $\xi$ involve only 3 different reciprocal lattice vectors $\bm{G} = \bm{0}, \xi\bm{G}_2, \xi\bm{G}_3$ for the top layer ($\bm{G}' = \bm{0}, \xi\bm{G}'_2, \xi\bm{G}'_3$ accordingly for the bottom layer). Thus, the effective inter-layer tunneling Hamiltonian can be written as:
\begin{eqnarray}
\mathcal{H}^{\xi}_{T, eff} &=& -\frac{1}{\Omega} \sum_{\bm{k}} \sum_{\alpha, \beta} c^{\dagger}_{\bm{k}, 1, \alpha} t_{\alpha\beta} (\xi\bm{K}) c_{\bm{k}, 2, \beta} + c^{\dagger}_{\bm{k}, 1, \alpha} t_{\alpha\beta} (\xi\bm{K} + \xi\bm{G}_2) e^{-i \xi\bm{G}'_2  \cdot \bm{d}_0}c_{\bm{k}+\xi\bm{G}_{M,2}, 2, \beta} \\\nonumber
&+&  c^{\dagger}_{\bm{k}, 1, \alpha} t_{\alpha\beta} (\xi\bm{K} + \xi\bm{G}_3) e^{-i \xi\bm{G}'_3  \cdot \bm{d}_0}c_{\bm{k}+\xi\bm{G}_{M,3}, 2, \beta} + h.c. \\\nonumber
&=& -\sum_{\bm{k}, \alpha\beta}  c^{\dagger}_{\bm{k}, 1, \alpha} w_{\alpha\beta} c_{\bm{k}, 2, \beta} + c^{\dagger}_{\bm{k}, 1, \alpha} w_{\alpha\beta} e^{-i \xi\bm{G}'_2  \cdot \bm{d}_0} c_{\bm{k}+\xi\bm{G}_{M,2}, 2, \beta} \\\nonumber
&+& c^{\dagger}_{\bm{k}, 1, \alpha} w_{\alpha\beta} e^{-i \xi\bm{G}'_3  \cdot \bm{d}_0}c_{\bm{k}+\xi\bm{G}_{M,3}, 2, \beta}  + h.c.,
\end{eqnarray}
where $w_{\alpha\beta} = t_{\alpha\beta}(K)/\Omega$, $ \bm{G}_{M,j} = \bm{G}_j - \bm{G}'_j$. Explicitly, the tunneling matrices connecting states at $\bm{k}$ on layer 1 to those at $\bm{k}, \bm{k} + \xi \bm{G}_{M,2}, \bm{k} + \xi \bm{G}_{M,3} $ on layer 2 are listed below:
\begin{eqnarray}
T_{\alpha\beta} (\bm{k}, \bm{k}) = -w_{\alpha\beta}, T_{\alpha\beta} (\bm{k}, \bm{k} + \xi \bm{G}_{M,2}) = -e^{-i \xi\bm{G}'_2  \cdot \bm{d}_0} w_{\alpha\beta}, T_{\alpha\beta} (\bm{k}, \bm{k} + \xi \bm{G}_{M,3}) = - e^{-i \xi\bm{G}'_3  \cdot \bm{d}_0} w_{\alpha\beta}.
\end{eqnarray}
For simplicity, we consider the case with $\bm{d}_0 = \bm{0}$ where metal sites from the two layers are aligned at zero twist angle, and assume $w_{\alpha\beta} = w \delta_{\alpha\beta}$ given that the inter-layer tunneling is usually dominated by spin-preserving processes. This leads to the form of the inter-layer tunneling Hamiltonian in Eq.3 of the main text.

\begin{figure}
\centering
\includegraphics[width=6in]{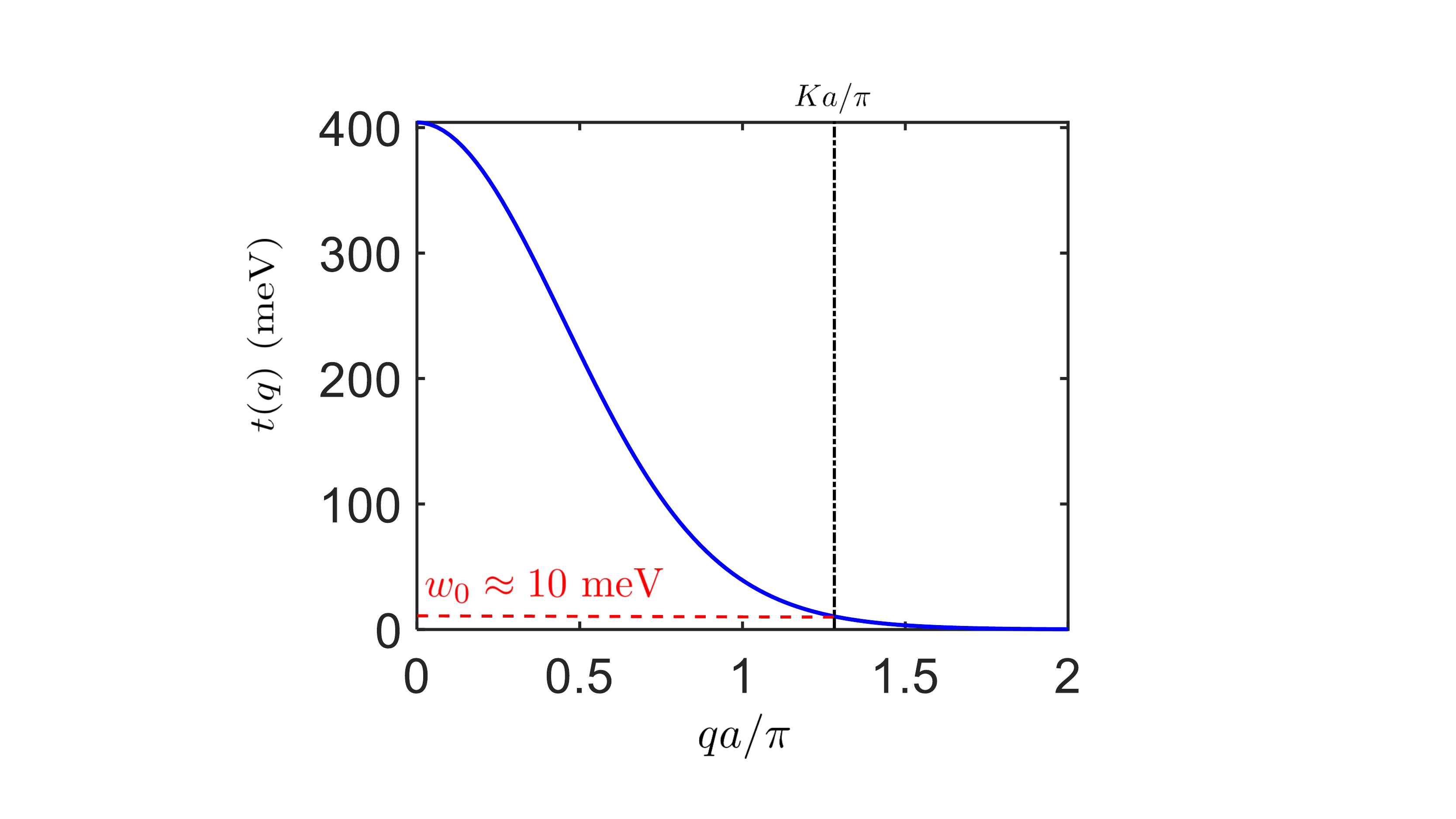}
\caption{Interlayer tunneling strength $t(q)$ as a function of $q$ calculated by numerical Fourier transform. Dashed vertical line in black indicates the value $w \approx 10$ meV extrapolated at $q = K$ where $Ka/\pi = 4/3$.}
\label{FIGS1}
\end{figure}

To evaluate the effective tunneling strength $t(\bm{q}=\bm{K})$ near the $\bm{K}$-points, we use the following simple model for $t(\bm{r})$ which incorporates the Slater-Koster $\sigma$-bonding between $d_{z^2}$-orbitals in the two different layers with an empirical exponential decay:
\begin{eqnarray}\label{eq:interlayertunneling}
t(\bm{r}) &=& [n^2(\bm{r}) - (l^2(\bm{r}) + m^2(\bm{r}))/2]^2 V_{dd\sigma}(\bm{r}), \\\nonumber
V_{dd\sigma}(\bm{r}) &=& t_0 e^{-\lambda \sqrt{d^2 + |\bm{r}|^2 } }.
\end{eqnarray}
Here, $d = 0.7$ nm is the inter-layer distance between two TMD monolayers \cite{SunS}, $n(\bm{r}), l(\bm{r}), m(\bm{r})$ are the directional cosines of the vector $\bm{r}$ connecting the two position vectors in the two layers, with
\begin{eqnarray}
n^2(\bm{r}) = \frac{d^2}{d^2 + |\bm{r}|^2  }, \hspace{1 cm} l^2(\bm{r}) + m^2(\bm{r})  =  \frac{|\bm{r}|^2}{d^2 + |\bm{r}|^2  }.
\end{eqnarray}

The value of $\lambda$ is estimated based on the empirical fact that $t' \approx 0.1 t$, where $t'$ and $t$ are the next-nearest-neighbor and neareast-neighbor hopping amplitudes. Thus, $t/t' = e^{\lambda a} \approx 10$, and $\lambda a = ln(10) \approx 2.3$. Note that the form of $t(\bm{r})$ in Eq.\ref{eq:interlayertunneling} is a radial function $t(\bm{r}) \equiv t(r)$, and its Fourier transform has the general form of:
\begin{eqnarray}
t(q) &=& \int_{0}^{\infty} dr r t(r) \int_{0}^{2\pi} d\theta e^{-i qr \cos \theta} \\\nonumber
               &=& 2\pi \int_{0}^{\infty} dr r J_0(qr) t(r),
\end{eqnarray}
where $J_0(x)$ is the Bessel function of order zero. With the form of $t(r)$ in Eq.\ref{eq:interlayertunneling}, $t(q)$ can be obtained by numerical integration with the only fitting parameter $t_0$. To fix $t_0$, we note that states near the $\Gamma$-point of the valence band states in MoS$_2$ also originate from $4d_{z^2}$-orbitals \cite{GuiBinS}, and the inter-layer coupling causes an energy splitting of $\Delta E \simeq 2 t(q=0) \approx 800$ meV at the $\Gamma$-point in the valence bands in aligned bilayer MoS$_2$ \cite{ZahidS}. This implies $t(q =0) \approx 400$ meV and allows us to fix $t_0$ and obtain the values of $t(q)$ for all $q$ as shown in Fig.\ref{FIGS1}. The effective interlayer coupling strength at the $K$-point in our continuum model is then extrapolated to be $w_0 \approx 10$ meV at $qa = Ka = 4\pi/3$ (dashed vertical line in black in Fig.\ref{FIGS1}). 
 
\section*{II. Chern numbers in moir\'{e} bands}

\subsection*{A. Numerical method for computing Chern numbers}

The Chern numbers of the CB moir\'{e} bands presented in Fig.2 and Fig.4 of the main text are calculated numerically using the standard formula introduced in Ref.\cite{SuzukiS}. The procedures of the numerical method is outlined below. For the sake of simplicity, we consider a given CB moir\'{e} band and drop the band index in the following.

First, we construct a discretized moir\'{e} Brillouin zone (MBZ) with an $N_x \times N_y$ grid. For each momentum point $\bm{k}$ in the grid, we evaluate the following complex quantities:
\begin{eqnarray}
U_x(\bm{k}) &\equiv& \braket{u(\bm{k} + \Delta k_x \hat{x} )|u(\bm{k})},\\\nonumber
U_y(\bm{k}) &\equiv& \braket{u(\bm{k} + \Delta k_y \hat{y} )|u(\bm{k})},
\end{eqnarray}
where $\ket{u(\bm{k})}$ is the periodic part of the Bloch state at $\bm{k}$, $\hat{x},\hat{y}$ are unit vectors in $k_x$ and $k_y$ directions, $\Delta k_x$ and $\Delta k_y$ measure the spacings between neighboring grid points in $k_x$ and $k_y$ directions. Note that the phase factors of $U_{x}$ and $U_{y}$ essentially account for the Berry connections on each link between two neighboring grid points. The gauge-invariant Berry flux threading through one plaquette around $\bm{k}$ is given by $\gamma_{\bm{k}} = \arg W(\bm{k})$, where $W(\bm{k})$ is the Wilson loop:
\begin{eqnarray}
W(\bm{k}) = U_{x} (\bm{k}) U_{y} (\bm{k} + \Delta k_x \hat{x}) U^{\ast}_{x} (\bm{k} + \Delta k_y \hat{y}) U^{\ast}_{y} (\bm{k}).
\end{eqnarray}The Chern number is then expressed as the sum over all Berry fluxes $\gamma_{\bm{k}}$ through a total number of $N_x \times N_y$ plaquettes:
\begin{eqnarray}
\mathcal{C} = \frac{1}{2\pi} \sum_{\bm{k}} \gamma_{\bm{k}}.
\end{eqnarray}
We note that to obtain accurate results for Chern numbers in the continuum model, the total number of grid points must be large enough such that $|\arg W(\bm{k})| < \pi$ for all $\bm{k}$, and the total number $M$ of the moir\'{e} Brillouin zones must also be large enough to ensure an accurate description of the low-energy physics at the exact $M \rightarrow \infty$ limit. In our model, we find that $N_x = N_y = 40$ and $M = 81$ produce accurate numerical results for $\mathcal{C}$ in the CB moir\'{e} bands.

\begin{figure}
\centering
\includegraphics[width=7in]{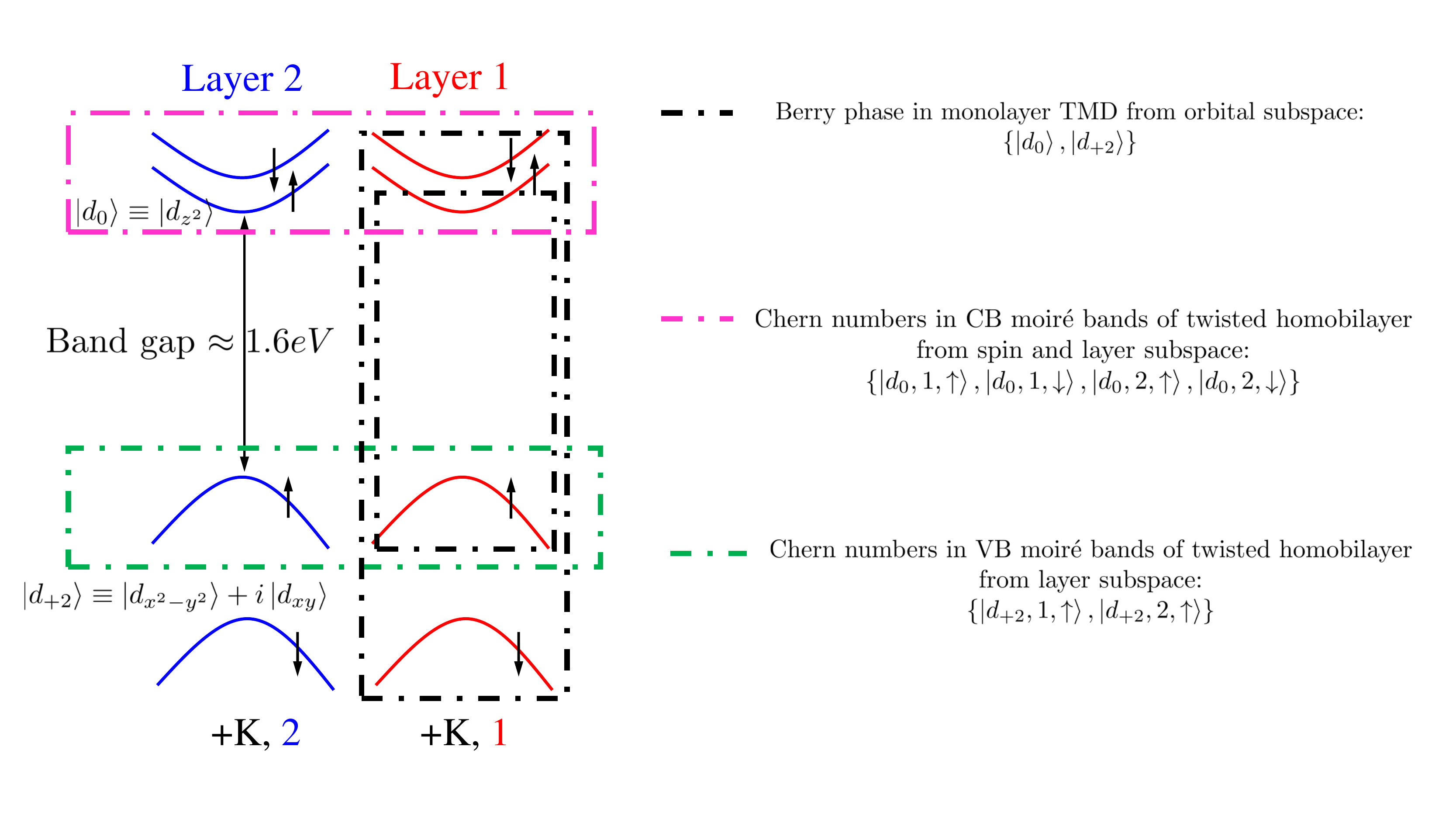}
\caption{Different sets of degrees of freedom involved in (i) Berry phase in monolayer TMD (encircled by black dashed rectangles), (ii) Chern numbers in CB moir\'{e} bands (encircled by pink dashed rectangles), and (iii) Chern numbers in VB moir\'{e} bands (encircled by green dashed rectangles).}
\label{FIGS2}
\end{figure}

\subsection*{B. Effects of Berry phase in monolayer transition-metal dichalcogenides}

It is known that a Berry phase with value close to $\pi$ can appear in a local region around $K$-points in the Brillouin zone of a monolayer transition-metal dichalcogenide (TMD) \cite{XiaoS}. This Berry phase originates from the massive Dirac Hamiltonian involving both the conduction and valence band edges at $+K$-point (see black dashed rectangles in Fig.\ref{FIGS2}), which is essentially different from both the Berry curvatures generated by the layer degrees of freedom in VB moir\'{e} bands \cite{WuS} (green dashed rectangles in Fig.\ref{FIGS2})) and the giant Berry curvatures (Fig.2(c) of the main text) generated by the spin and layer degrees of freedom in our continuum model  (pink dashed rectangles in Fig.\ref{FIGS2})).

While by summing over both the spin-up and spin-down valence bands in a monolayer TMD at one of the $K$-valleys, one acquires a total Berry phase close to $2\pi$ and appears to correspond to a “Chern number” of 1 for this valley,  such a “Chern number” is, strictly speaking, not well-defined. This is because the Berry curvatures in a monolayer are summed over two bands and integrated only over a local region around $+K$ in the (large) monolayer Brillouin zone. The integral is not guaranteed to be quantized by any topological reasons. In contrast, the Chern numbers in both the CB moiré bands in the main text and the VB moiré bands in Ref.\cite{WuS} are well-defined quantities because the Berry curvatures of each single band are integrated over the entire (small) moir\'{e} Brillouin zone. The resultant Chern numbers of each band from valley $\xi = +$ or $\xi = -$ are guaranteed to be quantized by topology. 

Importantly, we note that the Berry phase in monolayer TMDs do not affect the topology of CB moir\'{e} bands in a twisted homobilayer. This is because the Berry curvatures from the massive Dirac Hamiltonian is only on the order of $\sim 10 {\AA}^2$ \cite{XiaoS}, three orders of magnitude smaller than the Berry curvatures $\sim 10^4 {\AA}^2$ in the moir\'{e} Chern bands (Fig.2(c) of the main text) generated by spin and layer degrees of freedom. Thus, the topology of CB moiré bands is dictated by the spin and layer degrees of freedom, which justifies why the contributions from remote valence bands ($\sim$ 1.6 eV away from conduction band states) can be neglected.

\section*{III. Self-consistent Hartree-Fock calculations}
\subsection*{A. Hartree-Fock energy functional and Lagrange equations}

When the Chern bands in a twisted bilayer MoS$_2$ are half-filled, the low-energy effective Hamiltonian is written as:
\begin{equation}
\mathcal{H}_{eff} = \mathcal{H}_{0, eff} + \mathcal{H}_{I, eff}
\end{equation}
where the non-interacting Hamiltonian is given by
\begin{equation}
\mathcal{H}_{0, eff} = \sum_{\bm{k}} \sum_{\xi = \pm} E_{\xi} (\bm{k}) c^{\dagger}_{\xi} (\bm{k}) c_ {\xi} (\bm{k}) 
\end{equation}
Here, $\xi = \pm$ is the valley index, $E_{\xi}(\bm{k})$ is the energy of moir\'{e} band from valley $\xi$. The effective interacting Hamiltonian is given by:
\begin{eqnarray}
\mathcal{H}_{I, eff} &=& \sum_{\bm{k}, \bm{k}', \bm{q}} \frac{V(\bm{q})}{A} \sum_{\xi\xi'} \Lambda^{\xi}  (\bm{k}+\bm{q}, \bm{k}) \Lambda^{\xi'} (\bm{k}'-\bm{q} , \bm{k}') \\\nonumber
&\times &c^{\dagger}_{\xi}  (\bm{k}+\bm{q}) c^{\dagger}_{\xi'}  (\bm{k}'-\bm{q}) c_{\xi'}  (\bm{k}') c_{\xi}  (\bm{k}),
\end{eqnarray}
where $V(\bm{q}) = \int d^2 \bm{r} V(\bm{r}) e^{i\bm{q}\cdot\bm{r}}$ is the Fourier transform of the two-body interaction $V(\bm{r}_1 - \bm{r}_2)$, $A$ is the total area of the system, $\Lambda^{\xi}  (\bm{k}+\bm{q}, \bm{k}) \equiv \braket{u_{\xi, \bm{k}+\bm{q}} | u_{\xi, \bm{k}} }$ are the form factors arising from projecting the general density-density interaction to the active bands, which end up as the overlap between the cell-periodic parts of Bloch states within the same band from valley $\xi$. Note that due to the $U_{v}(1)$-symmetry, wave functions from different valleys do not overlap: $\Lambda^{+ -}  (\bm{k}', \bm{k}) \equiv \braket{u_{+, \bm{k}'} | u_{-, \bm{k}} }=0$.

As mentioned in the main text, the Coulomb energy scale $g_C \sim 10$ meV in twisted bilayer MoS$_2$ overhelms the band width $W \sim 1$ meV in the Chern bands found in the continuum model. This motivates us to consider possible ground states at $1/2$-filling as a general Hartree-Fock state of the form: $\ket{\Phi} = \prod _{\bm{k}} [w_{+}(\bm{k}) e^{i \theta_+ (\bm{k})} c^{\dagger}_{+} (\bm{k}) +  w_{-}(\bm{k}) e^{i \theta_- (\bm{k})} c^{\dagger}_{-} (\bm{k})]  \ket{\Phi_0}$, where $w_{\xi}(\bm{k})$ and $\theta_{\xi}(\bm{k})$ are real numbers and denote the amplitude and phase of the coefficients of valley $\xi$ at momentum $\bm{k}$. $\ket{\Phi_0}$ is the vaccuum state with all bands below the active bands being fully filled. The variational parameters can be determined by minimizing the ground state energy subject to the normalization condition $w^2_{+}(\bm{k}) + w^2_{-}(\bm{k}) = 1$. The energy of $\ket{\Phi}$ is  given by:
\begin{eqnarray}
E^{\Phi} \equiv \braket{\Phi|\mathcal{H}_{eff}|\Phi} = E^{\Phi}_0 + E_H + E^{\Phi}_F,
\end{eqnarray}
where $E^{\Phi}_0$, $E_H$ and $E^{\Phi}_F$ refer to the kinetic energy, Hartree energy, and the Fock exchange energy of $\ket{\Phi_0}$, respectively. It is straightforward to show that the Hartree term essentially accounts for the direct coupling: $E_{H} \propto n_{e}^2$, where $n_e$ is the electron density of the system, which is fixed at a given band filling. Thus, we neglect $E_H$ from now on and focus on $E^{\Phi}_0$ and $E^{\Phi}_F$:
\begin{eqnarray}
E^{\Phi}_0 &=& \sum_{\bm{k}} w^2_{+} (\bm{k}) E_{+}(\bm{k}) + w^2_{-} (\bm{k})E_{-}(\bm{k}), \\\nonumber
E^{\Phi}_F &=& -\sum_{\bm{k}, \bm{q}} \frac{V(\bm{q})}{A} \sum_{\xi\xi'} \Lambda^{\xi}  (\bm{k}+\bm{q}, \bm{k}) \Lambda^{\xi'} (\bm{k} , \bm{k}+\bm{q}) w_{\xi}(\bm{k}+\bm{q}) w_{\xi'}(\bm{k}+\bm{q}) w_{\xi}(\bm{k}) w_{\xi'}(\bm{k}) \\\nonumber
 &\times& e^{-i \theta_{\xi} (\bm{k}+\bm{q})} e^{i \theta_{\xi'} (\bm{k}+\bm{q})} e^{-i \theta_{\xi'} (\bm{k})} e^{i \theta_{\xi} (\bm{k})}.
\end{eqnarray}
Note that in order to minimize $E^{\Phi}_F$, the phase factors would adjust themselves such that $e^{-i \theta_{\xi} (\bm{k}+\bm{q})} e^{i \theta_{\xi} (\bm{k})} \Lambda^{\xi}  (\bm{k}+\bm{q}, \bm{k}) = |\Lambda^{\xi}  (\bm{k}+\bm{q}, \bm{k})|$. This reduces $E^{\Phi}_F $ to the following form:
\begin{eqnarray}
E^{\Phi}_F = -\sum_{\bm{k}, \bm{q}} \frac{V(\bm{q})}{A} \sum_{\xi\xi'} |\Lambda^{\xi}  (\bm{k}+\bm{q}, \bm{k})| |\Lambda^{\xi'} (\bm{k}+\bm{q}, \bm{k})| w_{\xi}(\bm{k}+\bm{q}) w_{\xi'}(\bm{k}+\bm{q}) w_{\xi}(\bm{k}) w_{\xi'}(\bm{k}) 
\end{eqnarray}
where we have used the identity $|\Lambda^{\xi'} (\bm{k}, \bm{k}+\bm{q})| = |\Lambda^{\xi'} (\bm{k}+\bm{q}, \bm{k})^{\ast}| = |\Lambda^{\xi'} (\bm{k}+\bm{q}, \bm{k})|$. To minimize $E^{\Phi}_0 + E^{\Phi}_F$ subject to the constraint $w^2_{+}(\bm{k}) + w^2_{-}(\bm{k}) = 1$, we introduce the Langrange multipliers $\lambda_{\bm{k}}$ such that 
\begin{equation}
\partial_{w_{\xi} (\bm{k})} \big[ E^{\Phi}_0 + E^{\Phi}_F  - \sum_{\bm{k}} \lambda_{\bm{k}} (w^2_{+}(\bm{k}) + w^2_{-}(\bm{k})) \big] = 0.
\end{equation}
This leads to the following two equations:
\begin{eqnarray} \label{eq:LagrangeEqn}
\begin{cases}
\big[E_{+}(\bm{k}) - \sum_{\bm{q}} g_{+}(\bm{q}, \bm{k}) w^2_{+}(\bm{k}+\bm{q})] w_{+}(\bm{k}) -[\sum_{\bm{q}} g_{+-}(\bm{q}, \bm{k})w_{+}(\bm{k}+\bm{q})) w_{-}(\bm{k}+\bm{q})\big] w_{-}(\bm{k})  = \lambda_{\bm{k}} w_{+}(\bm{k})\\
\big[E_{-}(\bm{k}) - \sum_{\bm{q}} g_{-}(\bm{q}, \bm{k}) w^2_{-}(\bm{k}+\bm{q})] w_{-}(\bm{k})   - [\sum_{\bm{q}}g_{+-}(\bm{q}, \bm{k}) w_{-}(\bm{k}+\bm{q}) w_{-}(\bm{k}+\bm{q})\big] w_{+}(\bm{k})  = \lambda_{\bm{k}} w_{-}(\bm{k})
\end{cases}
\end{eqnarray}
where we introduced the shorthand notation:
\begin{eqnarray}
\begin{cases}
g_{\xi}(\bm{q}, \bm{k}) = V(\bm{q}) |\Lambda^{\xi}  (\bm{k}+\bm{q}, \bm{k})|^2/A, (\xi = \pm) \\
g_{+-}(\bm{q}, \bm{k}) = V(\bm{q}) |\Lambda^{+}  (\bm{k}+\bm{q}, \bm{k})| |\Lambda^{-}  (\bm{k}+\bm{q}, \bm{k})| /A.
\end{cases}
\end{eqnarray}

As we seek for nontrivial solutions of Eq.\ref{eq:LagrangeEqn}, it is convenient to introduce the following two quantities:
\begin{eqnarray}\label{eq:definingeqn}
\Delta_{\xi}(\bm{k}) &=& \sum_{\bm{q}} g_{\xi}(\bm{q}, \bm{k}) w^2_{\xi}(\bm{k}+\bm{q}), \\\nonumber
\Delta_{+-}(\bm{k}) &=& \sum_{\bm{q}} g_{+-}(\bm{q}, \bm{k}) w_{+}(\bm{k}+\bm{q}) w_{-}(\bm{k}+\bm{q}),
\end{eqnarray}
which allows us to reformulate Eq.\ref{eq:LagrangeEqn} as an eigenvalue equation:
\begin{eqnarray}\label{eq:eigeneqn}
\begin{pmatrix}
E_{+}(\bm{k}) - \Delta_{+}(\bm{k})    &  -\Delta_{+-}(\bm{k})\\
-\Delta_{+-}(\bm{k})  &  E_{-}(\bm{k}) - \Delta_{-}(\bm{k}) 
\end{pmatrix}
\begin{pmatrix}
w_{+}(\bm{k}) \\
w_{-}(\bm{k})
\end{pmatrix}
= \lambda_{\bm{k}}
\begin{pmatrix}
w_{+}(\bm{k}) \\
w_{-}(\bm{k})
\end{pmatrix}.
\end{eqnarray}
Note that $\lambda_{\bm{k}}$ has the physical interpretation as the energy contribution from $\bm{k}$ to the total energy of the Hartree-Fock state. This motivates us to focus on the lower branch of eigenvalues of the $2 \times 2$ symmetric matrix:
\begin{eqnarray}\label{eq:LagrangeMultipiler1}
\lambda_1 (\bm{k}) =  \frac{1}{2} [E_{+}(\bm{k})+ E_{-}(\bm{k}) - (\Delta_{+}(\bm{k}) + \Delta_{-}(\bm{k})) ] - \sqrt{ \Delta^2_{+-}(\bm{k}) + \frac{1}{4} [E_{+}(\bm{k}) - E_{-}(\bm{k}) - (\Delta_{+}(\bm{k}) - \Delta_{-}(\bm{k})) ]^2}
\end{eqnarray}
with the corresponding eigenvector $[w_{+}(\bm{k}), w_{-}(\bm{k})]^T = [\cos(\phi_{\bm{k}}/2), \sin(\phi_{\bm{k}}/2)]^T$, where
\begin{eqnarray}
\sin(\phi_{\bm{k}}) &=& \frac{ \Delta_{+-}(\bm{k}) }{ \sqrt{ \Delta^2_{+-}(\bm{k}) + \frac{1}{4} [E_{+}(\bm{k}) - E_{-}(\bm{k}) - (\Delta_{+}(\bm{k}) - \Delta_{-}(\bm{k})) ]^2} } = 2 \cos(\phi_{\bm{k}}/2) \sin(\phi_{\bm{k}}/2), \\\nonumber
\cos(\phi_{\bm{k}}) &=& \frac{1}{2} \frac{ (\Delta_{+}(\bm{k}) - \Delta_{-}(\bm{k})) - (E_{+}(\bm{k}) - E_{-}(\bm{k}) )  }{ \sqrt{ \Delta^2_{+-}(\bm{k}) + \frac{1}{4} [E_{+}(\bm{k}) - E_{-}(\bm{k}) - (\Delta_{+}(\bm{k}) - \Delta_{-}(\bm{k})) ]^2} } = \cos^2(\phi_{\bm{k}}/2) - \sin^2(\phi_{\bm{k}}/2).
\end{eqnarray}

The defining equations (Eq.\ref{eq:definingeqn}) require that $\Delta_{+}(\bm{k}) , \Delta_{-}(\bm{k}), \Delta_{+-}(\bm{k}) $ satisfy the following self-consistent equations:
\begin{eqnarray}\label{eq:selfconsistenteqn2}
\Delta_{+}(\bm{k}) &=& \frac{1}{N} \sum_{\bm{k}'} J_{++}(\bm{k}', \bm{k}) \frac{1 + \cos(\phi_{\bm{k}'})}{2}, \\\nonumber
\Delta_{-}(\bm{k}) &=& \frac{1}{N}\sum_{\bm{k}'} J_{--}(\bm{k}', \bm{k}) \frac{1 - \cos(\phi_{\bm{k}'})}{2}, \\\nonumber
\Delta_{+-}(\bm{k}) &=& \frac{1}{N}\sum_{\bm{k}'} J_{+-}(\bm{k}', \bm{k}) \frac{\sin(\phi_{\bm{k}'})}{2}.
\end{eqnarray}
Here, $J_{\xi \xi'} (\bm{k}, \bm{k}')$ are the Fock exchange functions:
\begin{eqnarray}
J_{\xi \xi'}(\bm{k}', \bm{k}) = \Omega^{-1} \sum_{\bm{G}_M} V(\bm{k}' + \bm{G}_M - \bm{k}) |\Lambda^{\xi}(\bm{k}' + \bm{G}_M, \bm{k})| |\Lambda^{\xi'}(\bm{k}, \bm{k}' + \bm{G}_M)|,
\end{eqnarray}
where $\Omega$ is the area of the real-space moir\'{e} unit cell, $\bm{G}_M$ denotes the moir\'{e} reciprocal lattice vectors. 

\subsection*{B. Mean-field approximation}

To simplify Eq.\ref{eq:selfconsistenteqn2}, we employ the mean-field approximation which replaces the exchange coupling strengths $J_{\xi \xi'}(\bm{k}', \bm{k})$ by their averages: intravalley exchange coupling $J_0 \equiv \braket{  J_{+ +}(\bm{k}', \bm{k})} = \braket{  J_{--}(\bm{k}', \bm{k})} $, and the intervalley exchange coupling $J_1 \equiv \braket{  J_{+-}(\bm{k}', \bm{k})} $. This reduces Eq.\ref{eq:selfconsistenteqn2} to a coupled set of self-consistent equations involving only constant order parameters:
\begin{eqnarray}\label{eq:MFEqn}
\Delta_{+} &=& \frac{J_0}{N} \sum_{\bm{k}} \frac{1 + \cos(\phi_{\bm{k}})}{2}, \\\nonumber
\Delta_{-} &=& \frac{J_0}{N} \sum_{\bm{k}}  \frac{1 - \cos(\phi_{\bm{k}})}{2}, \\\nonumber
\Delta_{+-} &=& \frac{J_1}{N} \sum_{\bm{k}}  \frac{\sin(\phi_{\bm{k}})}{2},
\end{eqnarray}
where 
\begin{eqnarray}
\sin(\phi_{\bm{k}}) &=& \frac{ \Delta_{+-} }{ \sqrt{ \Delta^2_{+-} + \frac{1}{4} [E_{+}(\bm{k}) - E_{-}(\bm{k}) - (\Delta_{+} - \Delta_{-}) ]^2} } , \\\nonumber
\cos(\phi_{\bm{k}}) &=& \frac{1}{2} \frac{ (\Delta_{+} - \Delta_{-}) - (E_{+}(\bm{k}) - E_{-}(\bm{k}) )  }{ \sqrt{ \Delta^2_{+-} + \frac{1}{4} [E_{+}(\bm{k}) - E_{-}(\bm{k}) - (\Delta_{+} - \Delta_{-}) ]^2} }.
\end{eqnarray}

By adding and subtracting the first two equations in Eq.\ref{eq:MFEqn}, we end up with a compact form (Eq.5 of the main text):
\begin{eqnarray}\label{eq:MFEqn2}
\Delta_{+} + \Delta_{-} &=& J_0, \\\nonumber
\Delta_{+} - \Delta_{-} &=& \frac{J_0}{2N} \sum_{\bm{k}}  \frac{(\Delta_{+} - \Delta_{-}) - \delta(\bm{k})}{D(\bm{k})}, \\\nonumber
\Delta_{+-} &=& \frac{J_1}{2N} \sum_{\bm{k}}  \frac{\Delta_{+-}}{D(\bm{k})}
\end{eqnarray}
Here, $\delta(\bm{k}) = E_{+}(\bm{k}) - E_{-}(\bm{k})$, $D(\bm{k}) = \sqrt{\Delta^2_{+-} + \frac{1}{4} [\delta(\bm{k}) - (\Delta_{+} - \Delta_{-})]^2}$. Note that time-reversal enforces $\delta(-\bm{k}) = E_{+}(-\bm{k}) - E_{-}(-\bm{k}) = E_{-}(\bm{k}) - E_{+}(\bm{k}) = -\delta(\bm{k})$ to be an odd function of $\bm{k}$.

\subsection*{C. Hartree-Fock ground states under different regimes of mean-field exchange coupling $(J_0, J_1)$}

It is worth noting that there exist two possible solutions for the last equation in Eq.\ref{eq:MFEqn2}, which divides the whole set of solutions to Eq.\ref{eq:MFEqn2} into two different classes: (i) $\Delta_{+-} = 0$, and (ii) $\Delta_{+-} \neq 0$. Particularly, in case (i) where $\Delta_{+-} = 0$,  we have $D(\bm{k}) = \frac{1}{2} |\delta(\bm{k}) - (\Delta_{+} - \Delta_{-})|$, \textit{i.e.}, $\cos(\phi_{\bm{k}}) = \pm 1$ for all $\bm{k}$ (or equivalently, either $\Delta_{+} = J_0, \Delta_{-} = 0$ or $\Delta_{+} = 0, \Delta_{-} = J_0$). These two degenerate states are the \textbf{valley-polarized  (VP)} solutions which spontaneously break time-reversal symmetry. The energy of the two degenerate VP states is given by 
\begin{eqnarray}\label{eq:EVP}
E_{VP} = \sum_{\bm{k}} E_{+}(\bm{k}) - N J_0.
\end{eqnarray}

In case (ii) where $\Delta_{+-} \neq 0$, one can divide both sides of the $3^{\text{rd}}$ equation in Eq.\ref{eq:MFEqn2} by $\Delta_{+-}$, which simplifies the set of equations as below:
\begin{eqnarray}\label{eq:MFEqn3}
\Delta_{+} + \Delta_{-} &=& J_0, \\\nonumber
\Delta_{+} - \Delta_{-} &=& \frac{J_0}{J_1} (\Delta_{+} - \Delta_{-}) - \frac{J_0}{2N}\sum_{\bm{k}} \frac{\delta(\bm{k})}{D(\bm{k})}, \\\nonumber
1 &=& \frac{J_1}{2N} \sum_{\bm{k}}  \frac{1}{D(\bm{k})},
\end{eqnarray}
where the three unknown mean-field order parameters $\Delta_{+}, \Delta_{-}, \Delta_{+-}$ that characterize this state need to be solved self-consistently. In particular, we note that $\Delta_{+-} \neq 0$ implies a nontrivial \textbf{inter-valley order} associated with this Hatree-Fock state, in which the microscopic valley is no longer conserved (\textit{i.e.}, spontaneous breaking of valley-$U_{v}(1)$ symmetry). On the other hand, any solution with nonzero $\Delta_M \equiv \Delta_{+} - \Delta_{-}$ necessarily lifts the valley degeneracy and breaks time-reversal symmetry spontaneously. Thus, we call $\Delta_M$ as the \textbf{valley magnetic order}. As it turns out, under time-reversal symmetry (\textit{e.g.}, in the absence of any external magnetic fields), the only $U_{v}(1)$-breaking solution must satisfy the condition $\Delta_{+} = \Delta_{-}$, \textit{i.e.}, there is no valley magnetic order $\Delta_{M} = \Delta_{+} - \Delta_{-}$ involved. This solution is known as the (generalized) \textbf{inter-valley coherence (IVC)} state \cite{AdrianS} (note: $w_{+}(\bm{k}) = w_{-}(-\bm{k}) \neq 1/\sqrt{2}$ in general except at time-reversal-invariant momenta). According to Eq.\ref{eq:LagrangeMultipiler1}, the energy of the IVC state is:
\begin{eqnarray}
E_{IVC} &=& \sum_{\bm{k}}  \frac{1}{2} [E_{+}(\bm{k})+ E_{-}(\bm{k}) - (\Delta_{+} + \Delta_{-})] - D(\bm{k}) \\\nonumber
[E_{-}(\bm{k}) = E_{+}(-\bm{k}),  \Delta_{+} + \Delta_{-} = J_0]&=& \sum_{\bm{k}} [E_{+}(\bm{k}) - D(\bm{k})] - \frac{N}{2} J_0.
\end{eqnarray}
The ground state can then be determined by comparing the two energies $E_{IVC}$ and $E_{VP}$:
\begin{eqnarray}
\Delta E = E_{IVC} - E_{VP} = \frac{N}{2} J_0 - \sum_{\bm{k}} D(\bm{k}),
\end{eqnarray}
where $\Delta_{+}, \Delta_{-}, \Delta_{+-}$ involved in $D(\bm{k})$ are given by the self-consistent solutions of Eq.\ref{eq:MFEqn3}. The ground states obtained for each pair of $(J_0, J_1)$ map out a $J_0 - J_1$ mean-field phase diagram.

To develop some physical insights into the energetics of the system, we first consider some simple limits. First, we note that the flat band condition implies $J_0, J_1 \gg |\delta(\bm{k})|$. This combined with the self-consistency condition $1 = \frac{J_1}{2N} \sum_{\bm{k}} \frac{1}{D(\bm{k})}$ suggests $D(\bm{k})\sim J_1/2$. Thus, when the intravalley exchange coupling dominates over its intervalley counterpart: $J_0 \gg J_1$, we have $\Delta E \sim N(J_0 - J_1)/2 \gg 0$. In this case, the VP state is favored. By a similar argument, when the intervalley exchange coupling dominates: $J_1 \gg J_0$, $\Delta E \sim N(J_0 - J_1)/2 \ll 0$ and the IVC state always wins. In particular, in the entire $J_1 \geq J_0$ region, using the \textbf{harmonic mean-arithmetic mean (HM-AM) inequality}: 
\begin{eqnarray}
\frac{N}{\sum^{N}_{i=1} \frac{1}{x_i}} \leq \frac{\sum^{N}_{i=1} x_i}{N}, (x_i >0, \forall i = 1,...,N)
\end{eqnarray}
one can show that
\begin{eqnarray}
\sum_{\bm{k}} D(\bm{k}) \geq \frac{N^2}{\sum_{\bm{k}} \frac{1}{D(\bm{k})}} = \frac{N}{2} J_1,
\end{eqnarray}
where we used the last equation in Eq.\ref{eq:MFEqn3}. This shows that for $J_1 \geq J_0$, $\Delta E \leq N(J_0 - J_1)/2 \leq 0$ always holds, and the system favors the IVC state with nonzero $\Delta_{+-}$ and $\Delta_{M} = \Delta_{+} - \Delta_{-} = 0$.

Next, we consider the $J_0>J_1$ regime. As we discussed above, at $J_0 = J_1$, the system always favors the IVC state, while in the limit $J_0 \gg J_1$ the system should favor the VP state. This suggests that upon increasing the ratio $J_0/J_1$, a phase transition must happen somewhere in the $J_0>J_1$ regime. By solving the self-consistent equations (Eq.\ref{eq:MFEqn2}) numerically, we obtain the $J_0-J_1$ phase diagram in Fig.3 of the main text, which reveals that the VP state is favored almost immediately when the $J_0>J_1$ regime is accessed, and a first-order transition from IVC to VP phase occurs at $J_0 \simeq J_1$. This is due to the fact that the energetics of the system is essentially governed by $(J_0, J_1)$ under the condition $J_0, J_1 \gg |\delta(\bm{k})|$. As a result, $\Delta E \simeq N(J_0 - J_1)/2$ almost holds all the time, and the VP state is favored almost in the entire regime with $J_0 > J_1$.

\subsection*{D. Physically accessible regimes of $(J_0, J_1)$ under density-density repulsions}

We note that the discussions above are based on pure algebraic analysis --- it is clearly desirable to examine the accessible regimes of $(J_0, J_1)$ on physical grounds. Indeed, given a generic density-density repulsive interaction which respects $U_{v}(1)$ and $\mathcal{T}$, the relation between $J_0$ and $J_1$ can be derived. In particular, the intra-valley exchange coupling strength is
\begin{eqnarray}
J_0 &\equiv&  \braket{J_{++}(\bm{k}', \bm{k})} \\\nonumber
      & = &  \frac{1}{N^2 \Omega} \sum_{\bm{k}, \bm{k}'} \sum_{\bm{G}_M} V(\bm{k}' + \bm{G}_M - \bm{k}) |\Lambda^{+}(\bm{k}' + \bm{G}_M, \bm{k}) |^2 \\\nonumber
      & = & \frac{1}{N^2 \Omega} \sum_{\bm{k}, \bm{k}'} \sum_{\bm{G}_M} V(-\bm{k}' - \bm{G}_M + \bm{k}) |\Lambda^{-}(-\bm{k}, -\bm{k}' - \bm{G}_M) |^2 \\\nonumber
      & = &  \frac{1}{N^2 \Omega} \sum_{\bm{k}, \bm{k}'} \sum_{\bm{G}_M} V(\bm{k}' + \bm{G}_M - \bm{k}) |\Lambda^{-}(\bm{k}' + \bm{G}_M, \bm{k}) |^2 \\\nonumber
     & = & \braket{J_{--}(\bm{k}', \bm{k})}.
\end{eqnarray}
Note that in the equations above we used the fact that $V(\bm{q}) = V(-\bm{q})$ and the condition $\Lambda^{+}(\bm{k}' , \bm{k}) = \Lambda^{-}(-\bm{k} , -\bm{k}')$ imposed by time-reversal symmetry. For the inter-valley exchange coupling strength,
\begin{eqnarray}
J_1 &\equiv&  \braket{J_{+-}(\bm{k}', \bm{k})} \\\nonumber
      & = &  \frac{1}{N^2 \Omega} \sum_{\bm{k}, \bm{k}'} \sum_{\bm{G}_M} V(\bm{k}' + \bm{G}_M - \bm{k}) |\Lambda^{+}(\bm{k}' + \bm{G}_M, \bm{k})|| \Lambda^{-}(\bm{k}, \bm{k}' + \bm{G}_M) | \\\nonumber
     &\leq&  \frac{1}{N^2 \Omega} \sum_{\bm{k}, \bm{k}'} \sum_{\bm{G}_M} V(\bm{k}' + \bm{G}_M - \bm{k}) \frac{ |\Lambda^{+}(\bm{k}' + \bm{G}_M, \bm{k})|^2 + |\Lambda^{+}(-\bm{k}' - \bm{G}_M, -\bm{k})|^2}{2} \\\nonumber
   [V(\bm{q}) = V(-\bm{q})]  & = & \frac{1}{2N^2 \Omega} \big[ \big(\sum_{\bm{k}, \bm{k}'} \sum_{\bm{G}_M} V(\bm{k}' + \bm{G}_M - \bm{k}) |\Lambda^{+}(\bm{k}' + \bm{G}_M, \bm{k})|^2 \big) \\\nonumber
    & + & \big( \sum_{\bm{k}, \bm{k}'} \sum_{\bm{G}_M} V(-\bm{k}' - \bm{G}_M + \bm{k}) |\Lambda^{+}(-\bm{k}' - \bm{G}_M, -\bm{k})|^2 \big) \big] \\\nonumber
   & = & \frac{1}{2} 2 \braket{J_{++}(\bm{k}', \bm{k})} \\\nonumber
   & = & J_0.
\end{eqnarray}
Therefore, we identify the physically accessible points $(J_0, J_1)$ are located in the $J_0 \geq J_1$ regime of the phase diagram. Notably, the relation $J_0 \geq J_1$ derived above holds regardless of details of the wave functions and the microscopic interaction $V(\bm{q})$ as long as the interaction is repulsive in nature. Thus, as shown in Fig.3(a) of the main text, the VP state takes up a large portion of the physical regime with $J_0 \geq J_1$ except a narrow strip with $J_0 \approx J_1$ where the IVC state is favored. 

\subsection*{E. Case of general filling away from $1/2$}

Here, we briefly discuss the physics for general fillings away from $1/2$. For general fillings above $1/2$, one valley is fully filled, while the other is partially filled. On the other hand, for fillings below 1/2, one valley is empty while the other is partially filled. In both cases, the system is metallic and one expects anomalous Hall effects with non-quantized Hall conductance due to the large Berry curvatures hosted in the Chern bands (Fig.2(c) of the main text). 

Moreover, at full band filling (i.e., both valleys being fully filled), we expect time-reversal symmetry to be restored and the opposite Chern numbers in opposite valleys give rise to a pair of counter-propagating edge states. This effect can be termed as the \textbf{quantum valley Hall effect}, analogous to the quantum spin Hall effect in which opposite Chern numbers from opposite spin species give rise to a pair of counter-propagating edge states. 

\subsection*{F. Correlated topological insulating gap at $1/2$-filling}

As we mentioned in the Conclusion and Discussions section of the main text, under the flat band condition $W \ll g_C$ ($W$: band width, $g_C$: interaction strength) at small twist angles $\theta_0 \sim 1^\circ$, the actual topological insulating gap to be manifested experimentally at $1/2$-filling is not solely determined by the gap induced by Rashba SOCs in the non-interacting bands. Instead, it is largely governed by the correlation-induced insulating gap for the valley-polarized (VP) state, which is approximately given by the intra-valley exchange coupling strength $J_0 \sim g_C \approx 10$ meVs (see estimation of $g_C$ in the section “Correlated QAH state at $1/2$-filling” in the main text). 

\begin{figure}
\centering
\includegraphics[width=7in]{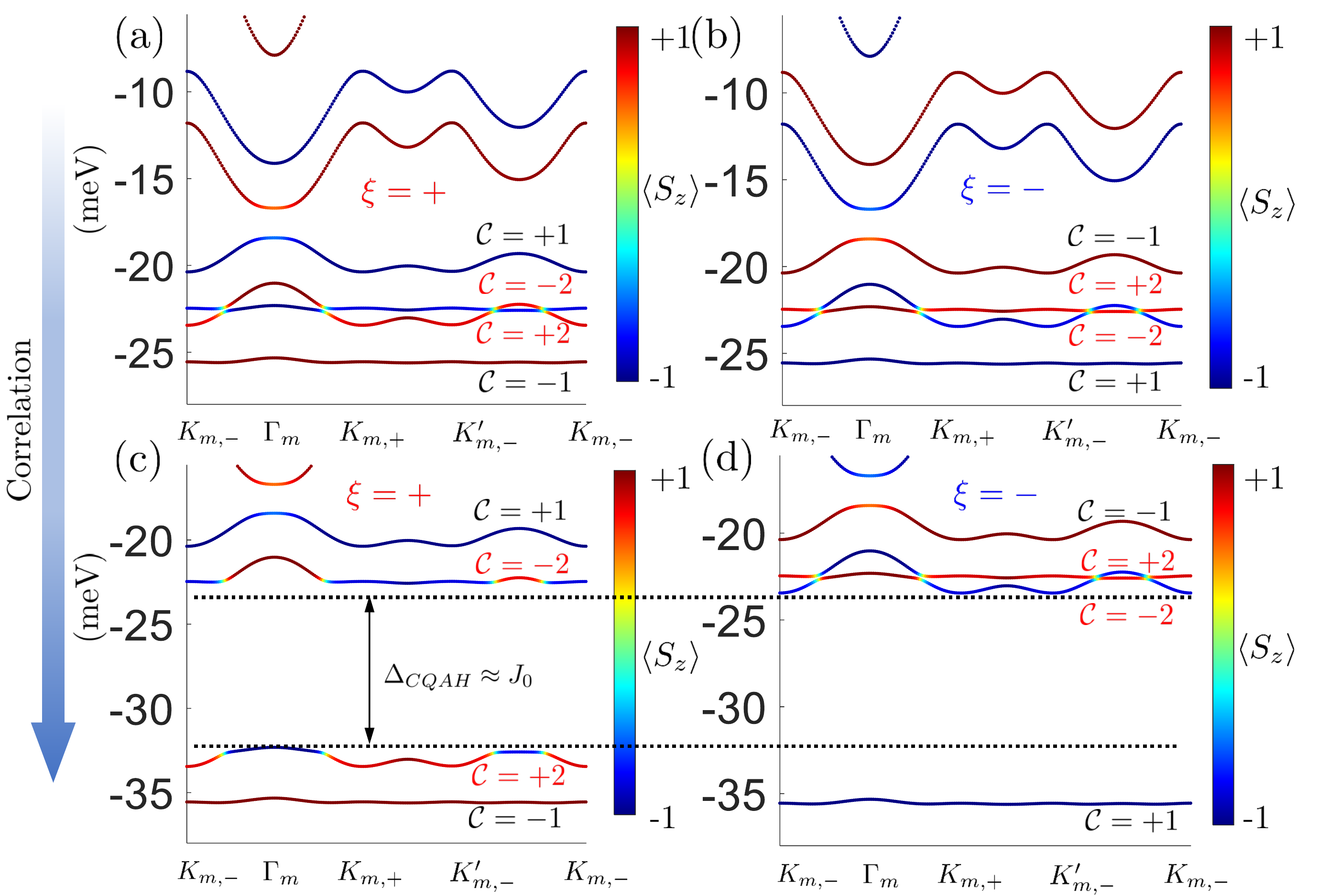}
\caption{(a)-(b) Non-interacting bands of twisted homobilayer MoS2 for (a) valley $\xi =+$, and (b) valley $\xi =-$ with Rashba strength $\alpha_{so} = 30$ meV$\cdot$ {\AA}. Dashed lines in (a)-(b) indicate location of the Fermi energy EF when the 2$^{\text{nd}}$ moiré bands are $1/2$-filled and interaction effects are neglected. In the absence of correlations, both bands from valley $\xi =+$ and valley $\xi =-$ are half-filled. (c)-(d) Correlated band structures when the 2$^{\text{nd}}$ moiré bands are half-filled, where the system enters the valley-polarized state. Electron correlations induce an insulating gap $\Delta_{CQAH} \sim J_0$, where we set $J_0=10$ meVs. 
}
\label{FIGS3}
\end{figure}

Physically, for a VP state, the intra-valley exchange interactions among electrons in the fully filled moiré band lowers the energy of each individual electron in this valley by an amount of $J_0$. On the other hand, the energy of the empty band in the other valley remains unaffected, as electron correlation cannot occur within an empty band. As a result, once the non-interacting bands acquire nonzero Chern numbers from any finite Rashba SOC, the correlated topological gap at $1/2$-filling would be on the order of $J_0 = 10$ meVs. To explicitly demonstrate this, we use our Hartree-Fock formalism (see Eq.\ref{eq:EVP}) to calculate the correlated band structures for both valley $\xi = +$ and valley $\xi = -$ when the pair of 2$^{\text{nd}}$ moir\'{e} bands are half-filled (Fig.\ref{FIGS3}). As the spontaneous $\mathcal{T}$-broken VP phase selects one out of the two valleys randomly, we assume without loss of generality that the 2$^{\text{nd}}$ moiré band from $\xi =+$ being fully filled, while its counterpart from $\xi =-$ is empty. Clearly, the topological insulating gap (indicated by double arrow in Fig.\ref{FIGS3}(c)-(d)) corresponds to the scale of $J_0$, instead of the gap induced by Rashba SOC in the non-interacting bands in Fig.\ref{FIGS3}(a)-(b). 

\section*{IV. Estimate of spatial variation in Rashba SOC strength}

As we discussed in the Conclusion and Discussion section in the main text, the formation of moir\'{e} pattern in principle leads to spatial variations of the Rashba SOC strength on the moir\'{e} scale (the amplitude of variations denoted by $\delta \alpha_{so}$ in the following). 

To estimate $\delta \alpha_{so}$, the key observation is that the spatial variation in Rashba strength is not independent of the moir\'{e} potential. In particular, the moir\'{e} potential causes a spatial variation in the inter-layer potential differences between the two constituent layers with its amplitude on the scale of $V_M \approx 10$ meVs (Table I of the main text). Given the inter-layer distance $d=7 {\AA}$ in bilayer MoS$_2$, this corresponds to a spatially modulating electric field with the amplitude of variations given by $\delta E_z = 1.4 mV\cdot {\AA}^{-1}$. Assuming a linear scaling relation $\delta \alpha_{so} = \lambda \delta E_z$, we obtain an estimated value of $\delta \alpha_{so} \approx 0.08 meV\cdot{\AA}$ by using a proportionality constant $\lambda = 0.0625 e\cdot {\AA}^2$ for MoS$_2$ extrapolated from Fig.7 of Refs.\cite{YaoS}. Thus, the amplitude of the spatial variation in Rashba SOC is negligibly small compared to previous reports on gate-induced Rashba SOC with $\alpha_{so} \sim 30 meV \cdot {\AA}$ in each layer \cite{BenjaminS}. This justifies the use of a uniform Rashba SOC in our continuum model as a good approximation.

\end{document}